%% file: main.tex
\documentclass[runningheads]{llncs}

\usepackage[ruled,vlined,linesnumbered]{algorithm2e}
\usepackage{subcaption}
\usepackage{float}
\usepackage{verbatim}
\usepackage{wrapfig}
\SetKw{KwBy}{by}
  \SetKwProg{Fn}{Function}{}{}
 \SetKwProg{Pro}{Proc}{}{}
 \SetKwFunction{FInit}{initialize}
  \SetKwFunction{Fwup}{warmup}
  \SetKwFunction{Fwforack}{wait\_for\_ack}
  \SetKwFunction{Fack}{ack}
  \SetKwFunction{Fbcast}{broadcast}
  \SetKwFunction{Fwtime}{wtime}
\SetKwFunction{Fmax}{max}
\SetKwFunction{Fssync}{start\_sync}
\SetKwFunction{Fesync}{stop\_sync}
\SetKwFunction{Fbarrier}{barrier}

\newcommand{\rememberlines}{\xdef\rememberedlines{\number\value{AlgoLine}}}
\newcommand{\resumenumbering}{\setcounter{AlgoLine}{\rememberedlines}}

\newcommand{\nonl}{\renewcommand{\nl}{\let\nl\oldnl}}% Remove line number for one line

\usepackage{tikz}
\usetikzlibrary{positioning}
\usetikzlibrary{decorations.pathreplacing,decorations.markings,decorations.pathmorphing}

\newcommand{\ignore}[1]{{}}
\newcommand{\skampi}{SKaMPI~}

\begin{document}
\title{SKaMPI-OpenSHMEM: Measuring OpenSHMEM Communication Routines}
\author{Camille Coti\inst{1,2}%\orcidID{0000-0002-1224-7786} 
\and
Allen D. Malony\inst{2} }
\authorrunning{C. Coti, A. Malony}
\institute{LIPN, CNRS UMR 7030, Université Sorbonne Paris Nord, France
\email{camille.coti@lipn.univ-paris13.fr}\\
\and
University of Oregon, USA\\
\email{malony@cs.uoregon.edu}}
\maketitle              % typeset the header of the contribution
\begin{abstract}
\input{abstract}
\keywords{First keyword  \and Second keyword \and Another keyword.}
\end{abstract}

\input{intro}
\input{related}
\input{measure}

\input{expe}

\input{conclu}

\bibliographystyle{splncs04}
\bibliography{skampi}

\end{document}

%% file: abstract.tex
  Benchmarking is an important challenge in HPC, in particular, to be
  able to tune the basic blocks of the software environment used by
  applications.  The communication library and distributed run-time
  environment are among the most critical ones.  In particular, many of
  the routines provided by communication libraries can be adjusted using
  parameters such as buffer sizes and
  % the
  communication algorithm.
  %used.
  As a consequence, being able to measure accurately the time taken by
  these routines is crucial
  % to be able
  in order to optimize them and achieve the
  best performance.  For instance, the SKaMPI library was designed to
  measure the time taken by MPI routines, relying on MPI's two-sided
  communication model to measure one-sided and two-sided peer-to-peer
  communication
  % routines
  and collective routines.
  In this paper, we discuss the benchmarking
  challenges specific to OpenSHMEM's communication model, mainly to
  avoid inter-call pipelining and overlapping when measuring the time
  taken by its routines.  We extend SKaMPI for OpenSHMEM for this
  purpose and demonstrate measurement algorithms that address
  OpenSHMEM's communication model in practice.  Scaling experiments are
  run on the Summit platform to compare different benchmarking
  approaches on the SKaMPI benchmark operations.  These show the
  advantages of our techniques for more accurate performance
  characterization.
  
%  We are going to release the full code on a public platform before the full paper submission.

%% file: intro.tex
\section{Introduction}
\label{sec:intro}

% \cite{hunold2016reproducible}
% \cite{hamid2006comparison}

% \url{https://www.github.com/coti/SKaMPI}

% For example, in MPI's non-blocking model, \cite{traff2007self} mentions that
% we can assume that  $MPI\_Send(n) \leq MPI\_Isend(n) + MPI\_wait$
% (a blocking communication is faster or at least as fast as the time
% to post a non-blocking communication and wait for its completion) but,
% but it also would make more sense to require
% $MPI\_Send(n) \approx MPI\_Isend(n) + MPI\_Wait$.

The ability to effectively utilize high-performance computing (HPC) systems
to their best potential depends heavily on tuned library implementations
specific to a machine's processor, memory, and communications components.
For distributed memory applications, the communication routines and
distributed runtime system should be implemented and optimized in close
association with the capabilities of the hardware interconnection network.
This poses special challenges for standard communication interfaces
designed to be portable across HPC platforms.  The performance
of low-level communication operations is important, but it is the
communication model semantics that ultimately defines the context for
correct execution.  Both aspects come into play when porting a communications
library from one HPC architecture to another.

Benchmarking is a powerful technique for understanding HPC performance.
When applied to the development and tuning of scalable distributed systems,
especially portable parallel communication libraries, benchmarking
can provide valuable insight for identifying high-value settings
of parameters and algorithm variants for different use scenarios.
The design of a benchmarking methodology and framework that can
elaborate communication model behaviors and
correctly generate test cases is highly relevant for achieving
productive outcomes.  It serves to maintain a coherent plan for
measurement and analysis during the performance characterization
and library tuning process.

The original goal of the research reported in this paper was to develop
a benchmarking system for OpenSHMEM that could be use to tune
OpenSHMEM implementations across multiple HPC machines.
Most important to our work was designing a benchmarking methodology
that was consistent with the OpenSHMEM standard and systematic in its
processing.  Unfortunately, only OpenSHMEM mini-benchmarks existed at the
time the research began.  While we anticipated that we would have to
develop the tests for most of the OpenSHMEM operations, we wondered
if we could reuse the high-level structure and methods of the SKaMPI
benchmarking system~\cite{reussner1998skampi}.
The paper reports our experience and success
in following this strategy.

There are four research contributions deriving from our work.  First,
we produced a "first of its kind" fully functional benchmarking system
for OpenSHMEM, based on the \skampi methodology and framework.
Second, we show how the \skampi methodology and framework could be
reused for OpenSHMEM purposes.  This outcome could be beneficial to
extending \skampi with other communication libraries in the future.
Third, we describe how the tests of communication 
routines specific to the OpenSHMEM standard are constructed.
Finally, we demonstrate our benchmarking system on the Summit
platform and report detailed analysis results.

The rest of the paper is structured as follows.
In Section~\S\ref{sec:related} we discuss related research work in
the performance measurement, analysis, and benchmarking of
communication libraries.  Here we introduce the former \skampi
work for MPI.  Section~\S\ref{sec:measure} looks at the specific
problem of measuring OpenSHMEM routines and the challenges of
creating a portable benchmarking solution based on \skampi
for characterization and tuning.
Our experimental evaluation is presented in
in Section~\S\ref{sec:expe}.  We show the use of our solution
on the DOE Summit system at Oak Ridge National Laboratory (ORNL).
Finally, we conclude and describe future directions.

%% file: related.tex
\section{Related works}
\label{sec:related}

% General stuff about communication performance measurement

Measuring the time taken by communications can be performed in two contexts.
It can be made on parallel applications, in order to determine how much time
the application spends communicating.
% Such systems are used for
Various robust \emph{profiling} and \emph{tracing} systems can be used,
% and we can cite, among others,
such as TAU \cite{shende2006tau}, VTune \cite{reinders2005vtune},
% and
Scalasca \cite{geimer2010scalasca}, Score-P \cite{knupfer2012score},
and EZTrace \cite{trahay2011eztrace}.
The objective is to characterize communication performance for the routines
actually used by the application.

The other context has to do with analyzing communications performance
for the purpose of tuning
% in which the time spent in communications can be measured is to tune
the communication routines (point-to-point \cite{brightwell2003evaluation}
or collective routines \cite{luo2020han,hunold2018autotuning}),
and to make sure that they fulfill performance
requirements \cite{traff2007self} on the system of interest.
This \textit{benchmarking} approach is fundamentally different from above,
but can lead to important outcomes that contribute to better application
communication performance.
The \textit{Special Karlsruhe MPI benchmark (SKaMPI)}
was created to benchmark MPI communications for supercomputer
users and system administrators who want to tune their
MPI libraries \cite{lastovetsky2008mpiblib},
evaluate and chose algorithms for
collective communications \cite{worsch2002benchmarking,worsch2003benchmarking},
and ensure performance portability of the MPI library across
platforms \cite{reussner2001achieving}.

% How we measure

Measuring the time spent in communications
% looks into
should consider parameters that might potentially be exploited by applications.
Regardless of the communication library used, the \emph{effective bandwidth}
can be measured
% specifically by
directly \cite{rabenseifner2001parallel}.
% More specifically,
However, each library will have communication routines that
will need specific measurement techniques to understand their
operation.
For instance,
% how can
peer-to-peer communications are interesting because certain libraries
might allow overlap with computation.
The measurement methodology to capture phenomena in such cases
% might
can be non-trivial.
Moreover, designing benchmarking methods that can be applied
across communication libraries is a challenge.

\skampi overcomes most limitations of previous benchmarks such as
PARKBENCH \cite{hey2000development} and \emph{mpbench} \cite{mucci1998mpbench},
as described in \cite{reussner1998skampi}.
For instance, \emph{mpbench} reduces the number of calls to the timer
by calling the measured function several times in a loop and measuring
the total time taken by the loop.
However, some pipelining
% can establish
might occur between consecutive calls;
for instance, when measuring tree-based collective operations,
or point-to-point communications that do not ensure remote completion when
the sender call returns.

In order to eliminate obvious pipelining between some collective operation calls,
\emph{mpbench} uses a different root for the operation
(for broadcasts and reductions) at every iteration.
However, depending on the communication topology used,
this might not be enough and in some cases a pipeline can
still establish between consecutive calls.
Other algorithms have been designed to eliminate inter-call
pipelining and perform an accurate measurement of these calls,
relying on the synchronizing model of MPI peer-to-peer
communications to enforce separation of consecutive
collective routines \cite{SK99}, and on the synchronization
model of MPI2 one-sided communications \cite{augustin2005benchmarking}.

% A bit of theory

Theoretical models are a close relative to
benchmarking and can complement its objectives,
in that they utilize empirical values measured
on target machines.
The representative \emph{LogP} model \cite{culler1993logp}
expresses point-to-point communications using four parameters:
the \emph{send and receive overheads}, which are the time to
prepare the data to send it over the network and the time to get
it from the network and provide it to the application
(denoted $o_s$ and $o_r$), the \emph{wire latency}, which is the
time for the data to actually travel through the network
(denoted $L$), and the \emph{gap}, which is the the minimum interval
between consecutive communications (denoted $g$).
Since $o$ and $g$ can overlap, the LogP model encourages overlapping with
computation with communication.
% enough?

%% file: measure.tex
\section{Measuring OpenSHMEM communication routines}
\label{sec:measure}

SKaMPI's measuring infrastructure and synchronization algorithms are described and evaluated in \cite{hunold2015mpi}.
Our objective is to utilize the \skampi framework for benchmarking OpenSHMEM communication.
However, considering studies of potential clock drift \cite{hunold2015impact},
% From these works,
we know that both barrier and window-based process synchronization suffer from drift and the processes might lose their synchronization as the measurement progresses.
SKaMPI's window-based clock synchronization can measure operations very accurately, but the logical global clocks drift quickly, so only a small number of MPI operations can be measured precisely. The hierarchical algorithm presented in \cite{hunold2015impact} has a smaller clock drift, but the processes still skew during the measurement. As a consequence, we cannot rely on this synchronization only to perform our measurements, and whenever possible, we need to design measurement strategies that rely on more precise measurements than just a synchronization followed by a call to the measured routine.

Scalable clock synchronization algorithms are presented in \cite{hoefler2010accurately} and can achieve synchronization in $O(log(P))$ rounds, whereas SKaMPI's algorithm takes $O(p)$ rounds.
% This question
Adopting a better algorithm is related to the \emph{infrastructure} that supports the measurements and is out of the scope of this paper.
However, it is important
% to note
that we
% need to
use lightweight timing mechanisms that are non-perturbing
relative to the granularity of the artifact being measured.  In some
fine-grained measurement cases, we had
to update \skampi timing methods.

% Very good paper
% http://hunoldscience.net/paper/mpi_clocksync_sahu_2015.pdf

\input{measurep2p}

\input{measurecoll}
\input{measureothers}

%% file: measurep2p.tex
\subsection{Point-to-point communication routines}
\label{sec:measure:p2p}

\paragraph{Blocking operations.}
OpenSHMEM includes two categories of blocking point-to-point communication routines: remote memory access routines and atomic operations. In these two categories, we have two types of routines: those that return as soon as possible and not when the data has actually been delivered, and those that return when the data has been delivered in the destination buffer.

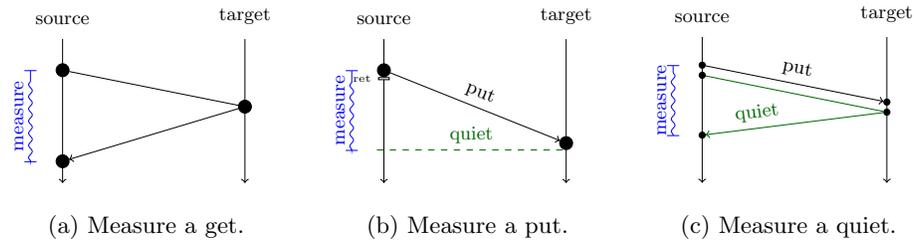
\begin{figure}
\vspace*{-.2in}
\centering
\begin{subfigure}[b]{0.3\textwidth}
\resizebox{\linewidth}{!}{\input{figures/get}}
\caption{\label{fig:measure:p2p:get}Measure a get.}
\end{subfigure}
\hfill
\begin{subfigure}[b]{0.3\textwidth}
\resizebox{\linewidth}{!}{\input{figures/put}}
\caption{\label{fig:measure:p2p:put}Measure a put.}
\end{subfigure}
\hfill
\begin{subfigure}[b]{0.3\textwidth}
\resizebox{\linewidth}{!}{\input{figures/quiet}}
\caption{\label{fig:measure:p2p:quiet}Measure a quiet.}
\end{subfigure}
\caption{\label{fig:measure:p2p}Different measurement cases for blocking operations.}
\vspace*{-.4in}
\end{figure}

Routines from the latter category can be called \emph{fetching} or \emph{get-based}. Their completion time corresponds to the time elapsed between the call and the return of the communication routine (see Figure \ref{fig:measure:p2p:get}),
% so they are
making them trivial to measure.
Routines from the former category can be called \emph{non-fetching} or \emph{put-based}. They are supposed to return as soon as the source buffer can be reused and not when the data has actually been delivered to the destination buffer. Since OpenSHMEM is a one-sided communication model, the target process does not participate in the communication: a global synchronization routine like a barrier cannot ensure completion of the operation, since the target process can enter the barrier and exit before the one-sided operation completes (and, since the order of operations is not guaranteed, before it has even reached the target process). 

Completion of the operation can be ensured with {\tt shmem\_quiet}. Therefore, we want to measure the time elapsed between the call to the communication routine and the return of {\tt shmem\_quiet} (see Figure \ref{fig:measure:p2p:put}). However, calling {\tt shmem\_quiet} can have a cost. Therefore, we need to measure the cost of an almost empty call to {\tt shmem\_quiet} and with substract it
% to
from the measured time. 

The {\tt shmem\_quiet} routine ensures completion of the outgoing put-based operations.
It works on all active communications with remote processes for the
calling process.  To measure the routine, therefore,
we must issue a {\tt shmem\_put} for one byte before the {\tt shmem\_quiet}.
We cannot exclude this {\tt shmem\_put} from the measurement, because some
communication engines might make it progress and complete before they schedule
the communication with all the remote processes involved by {\tt shmem\_quiet}.
However, it sends a small message and its latency should be combined with the
latency of the first part of the {\tt shmem\_quiet}, so this {\tt shmem\_put}
should have a negligible impact on the measurement of {\tt shmem\_quiet}
(see Figure \ref{fig:measure:p2p:quiet}).

Some implementations and some networks implement {\tt shmem\_put} as a non-blocking communication that only ensures that the sending buffer on the source process is reusable after the call exits (as in OpenMPI\footnote{SHA 62362849cae65b2445723e426affc2bb7918a6c8} \cite{shamis2015ucx}).
% , while
Others implement it like a blocking communication
(oshmpi\footnote{SHA 776449f6ea0368b61450b0c37e83463357f6f1bf} implements it as two {\tt MPI\_Isend} followed by a {\tt MPI\_Waitall} \cite{ghosh2014openshmem,osti_1771788}). Hence, measuring the time spent in the call to {\tt shmem\_put} is relevant.
% , and
We
% are providing
provide a function for this.

% TODO function names

\paragraph{Non-blocking operations.}
% TODO
With some implementations and some networks, the call to the communication routine can just post the communication and the communication is performed in {\tt shmem\_quiet}. Therefore, we can measure the time taken by this call to {\tt shmem\_quiet} for a given communication size. In the preamble of the measurement routine, we measure the time to perform a complete non-blocking communication (including the call to {\tt shmem\_quiet}) for the same buffer size. Then we post the operation and wait twice the measured time. Then we measure the time spent in {\tt shmem\_quiet}. If the library has good overlap capabilities, the communication will be performed during the wait, and the call to {\tt shmem\_quiet} will not do anything. Otherwise, the communication will be performed
% on
in {\tt shmem\_quiet}. 

Hence, we are providing four functions to measure non-blocking \emph{put} operations and four functions to measure non-blocking \emph{get} operations:
\begin{itemize}
    \item \verb|Shmem_{Put,Get}_Nonblocking_Full|
    measures the full completion of a non-blocking operation, with a \verb|shmem_put_nbi| or a \verb|shmem_get_nbi| immediately followed by a \verb|shmem_quiet|.
    \item \verb|Shmem_{Put,Get}_Nonblocking_Quiet| measures the time spent in\\
    \verb|shmem_quiet|, called immediately after \verb|shmem_put_nbi| or \verb|shmem_get_nbi|. If there is no overlap, most of the communication time is expected to be spent here. Otherwise, this call should be fast.
    \item \verb|Shmem_{Put,Get}_Nonblocking_Post| measures the time spent in the call to \verb|shmem_put_nbi| or \verb|shmem_get_nbi|. This call should be fast and the communication should not be performed here, otherwise the communication cannot be overlapped with computation.
    \item \verb|Shmem_{Put,Get}_Nonblocking_Overlap| measures the time spent in\\ \verb|shmem_put_nbi| or \verb|shmem_get_nbi| and the time spent in \verb|shmem_quiet|, separated by a computation operation that should take about twice the time to completed the full (non-blocking) communication.
\end{itemize}

These routines can be used to evaluate the overlapping capabilities of the library, by showing how much time is spent posting the non-blocking communications, waiting for them to complete, and comparing the time spent in these routines when they are and are not separated by a computation.

%% file: figures/get.tex
\begin{tikzpicture}

\node (sb) at ( 0, 0 )[label={[label distance=-1mm]above:source}]{};
\node (se) at ( 0, -2.6 ){};
\node (tb) at ( 3, 0 )[label={[label distance=-1mm]above:target}]{};
\node (te) at ( 3, -2.6 ){};

\draw[->,draw] (sb) -- (se);
\draw[->] (tb) -- (te);

\node(A) [below=4mm of sb,circle,inner sep=0pt,draw, fill]{o};
\node(B) [below=1cm of tb,circle,inner sep=0pt,draw, fill]{o};
\node(C) [below=1.9cm of sb,circle,inner sep=0pt,draw, fill]{o};;

\draw[->] (A) -- (B) -- (C);

\node (begin)[left=3mm of A] {};
\node (end)[left=3mm of C] {};

\draw [|-|, color=blue,decorate,decoration={snake,amplitude=.3mm,segment length=2mm,post length=1mm}](end.center) -- (begin.center)
node [midway,above,sloped] {measure};

\end{tikzpicture}

%% file: figures/put.tex
\begin{tikzpicture}

\node (sb) at ( 0, 0 )[label={[label distance=-1mm]above:source}]{};
\node (se) at ( 0, -2.6 ){};
\node (tb) at ( 3, 0 )[label={[label distance=-1mm]above:target}]{};
\node (te) at ( 3, -2.6){};

\draw[->,draw] (sb) -- (se);
\draw[->] (tb) -- (te);

\node(A) [below=4mm of sb,circle,inner sep=0pt,draw, fill]{o};
\node(B) [below=1.6cm of tb,circle,inner sep=0pt,draw, fill]{o};
%\node(C) [below=3cm of sb,circle,inner sep=0pt,draw, fill]{o};;
\draw[->] (A) -- (B) node [midway,above,sloped] {put};

\node[rectangle,below=0mm of A,draw=black, fill=white, inner sep=0pt,minimum width=5pt] (ret)  {};
\node[left=0 mm of ret] {\tiny ret};

\node (begin)[left=3mm of A] {};
\node (C) at  (A |- B.south) {};
\node (end)[left=3mm of C] {};

\draw[color=black!60!green,dashed](C.west) -- (B.south) node [midway,above] {quiet};

\draw [|-|, color=blue,decorate,decoration={snake,amplitude=.3mm,segment length=2mm,post length=1mm}](end.center) -- (begin.center)
node [midway,above,sloped] {measure};

\end{tikzpicture}

%% file: figures/quiet.tex
\begin{tikzpicture}

\node (sb) at ( 0, 0 )[label={[label distance=-1mm]above:source}]{};
\node (se) at ( 0, -2.6 ){};
\node (tb) at ( 3, 0 )[label={[label distance=-1mm]above:target}]{};
\node (te) at ( 3, -2.6){};

\draw[->,draw] (sb) -- (se);
\draw[->] (tb) -- (te);

\node(A) [below=4mm of sb,circle,inner sep=0pt,draw, fill,minimum size=1mm]{};
\node(B) [below=1cm of tb,circle,inner sep=0pt,draw, fill,minimum size=1mm]{};

\node(C) [below=.5mm of A,circle,inner sep=0pt,draw, fill,minimum size=1mm]{};
\node(D) [below=.5mm of B,circle,inner sep=0pt,draw, fill,minimum size=1mm]{};

\node (E)  [circle] at  (C |- D.south) {};
\node(F) [below=1mm of E,circle,inner sep=0pt,draw, fill,minimum size=1mm]{};

\draw[->] (A) -- (B) node [midway,above,sloped] {put};
\draw[->, color=black!60!green] (C) -- (D) -- (F) node [pos=.7,above,sloped] {quiet};

\node (begin)[left=3mm of A] {};
\node (end)[left=3mm of F] {};

\draw [|-|, color=blue,decorate,decoration={snake,amplitude=.3mm,segment length=2mm,post length=1mm}](end.center) -- (begin.center)
node [midway,above,sloped] {measure};

\end{tikzpicture}

%% file: measurecoll.tex
% \subsection{Measuring collective operations}
\subsection{Collective operations}
\label{sec:measure:coll}

\paragraph{Broadcast.}

\begin{wrapfigure}[14]{r}{0.56\textwidth}
%\begin{figure}[htt!]
\vspace*{-.35in}
\begin{algorithm}[H]\small
   $t\_bcast \gets 0.0$\;
   \Fbarrier{}\;
   \For{$i\gets0$ \KwTo $iterations$ \KwBy $1$}{
	 $t1 \gets$ \Fwtime{}\;
      \Fbcast{ $buffer, root$ }\;
   \Fbarrier{}\;
	 $t2 \gets$ \Fwtime{}\;
	 $\mathrm{t\_bcast} \gets \mathrm{t\_bcast} + (t2 - t1)$\;
    }
 $t\_bcast \gets t\_bcast / iterations$\;
 $t\_barrier  \gets \mathrm{time\_barrier}()$\;
 $t\_bcast \gets t\_bcast - t\_barrier$\;
 \caption{Barrier-synchronized broadcast measurement.\label{algo:measure:coll:bcastbarrier}}
\end{algorithm}
%\end{figure}
\vspace*{-.1in}
\end{wrapfigure}
When measuring a broadcast, a major challenge concerns how to avoid
% consists in avoiding
a pipeline that might occur
% to establish
between consecutive communications.
Therefore,
% when we measure a broadcast,
we want to separate consecutive broadcasts,
while avoiding external communication costs to be included. 
An initial possibility consists in separating consecutive broadcast with a barrier, and subtracting the time to perform a barrier (measured separately), as show by algorithm \ref{algo:measure:coll:bcastbarrier}.

\begin{figure}[htt!]
\vspace*{-.2in}
\centering
    \resizebox{.7\linewidth}{!}{\input{figures/bcast_barrier}}
\vspace*{-.1in}    \caption{Barrier-synchronized broadcast.}
    \label{fig:measure:coll:bcastbarrier}
\vspace*{-.2in}    
\end{figure}
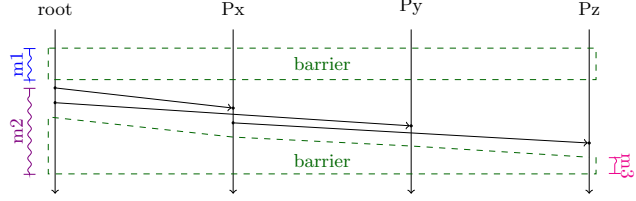

However, since with some broadcast algorithms,
% some
processes might exit early, the barrier performed after the
broadcast could become
% might be very
unbalanced and take a time different than a barrier performed by a set of already more or less synchronized processes. For instance, Figure \ref{fig:measure:coll:bcastbarrier} depicts a case when a process ends after the other ones, so the barrier can be finalized as soon as it enters it and all the other processes are already synchronized ($m3 < m1$). 
Moreover, a barrier does not truly synchronize the processes.
For instance, they can be implemented as a reduction followed by a broadcast using the same root, and depending on which process is the root of these operations, the end of the second barrier in Figure \ref{fig:measure:coll:bcastbarrier} can create a skew between processes.

\begin{wrapfigure}[17]{r}{0.56\textwidth}
%\begin{figure}[htt!]
\vspace*{-.35in}
 \begin{algorithm}[H]\small
  $root \gets 0$\;
  \For{ $i \gets 0$ \KwTo $iterations$ \KwBy $1$} {
   $t1 \gets$ \Fssync{}\;
   \Fbcast{ $buffer, root$ }\;
   $t2 \gets$ \Fesync{}\;
   $t\_bcast \gets t\_bcast + (t2 - t1)$\;
}
 $t\_bcast \gets t\_bcast / iterations$\;
 \caption{
 \begin{flushleft}
 Active synchronization-based.\label{algo:bcastsynchro}
 \end{flushleft}
 }
\end{algorithm}
\begin{algorithm}[H]\small
   $t1 \gets$  \Fssync{}\;
\For{ $root \gets 0$ \KwTo $size$  \KwBy $1$ }{
      \Fbcast{ $buffer, root$ }\;
}
$e\_time \gets$ \Fesync{}\;
$t\_bcast \gets (e\_time - s\_time)/size$\;
 \caption{In rounds.\label{algo:bcastrounds}}
\end{algorithm}
%\end{figure}
\end{wrapfigure}
SKaMPI provides a time-based synchronization, provided by {\tt start\_synchronization} and {\tt stop\_synchronization} routines. We can measure the broadcast operation as in algorithm \ref{algo:bcastsynchro}.
% TODO say more here about the synchro

Another possibility to try to avoid overlap and get the full
% extend
extent of a broadcast
% consists in
by doing broadcasts in rounds.
Each process is
% , at its time,
made the root of a broadcast and multiple broadcasts are performed.
However, depending on the broadcast topology, there might still be some overlap between consecutive broadcasts.
Algorithm~\ref{algo:bcastrounds}
was introduced by \cite{SK99}. In turns, processes acknowledge completion of their part of the broadcast to the root. The time to perform an acknowledgment is measured prior to the measurement, and subtracted from the total time. 

\begin{wrapfigure}[14]{r}{0.56\textwidth}
%\begin{figure}[htt!]
\vspace*{-9mm}   
\begin{algorithm}[H]
\small
  \Fn{\FInit{$task$}}{
      \tcc{Measure ack}
      $t1 \gets $ \Fwtime{}\;  
      \If{ $root == rank$ }{ \label{algo:SK:bcastinit:begin}
    	\Fack{task}\;
    	\Fwforack{}\;
      } \ElseIf{ $task == rank$ } {
    	\Fwforack{}\;
    	\Fack{root}\;\label{algo:SK:bcastinit:end}
      }
      $rt1 \gets$ \Fwtime{} $- t1$\;  
  \KwRet $rt1$\;
}
\rememberlines
\caption{(Part 1) One-sided communications broadcast measurement.\label{algo:SK}}
\end{algorithm}
%\end{figure}
\end{wrapfigure}
However, this algorithm was designed in a two-sided communication model, with synchronous communications. Conveniently, OpenSHMEM provides a routine that waits until a variable located in the shared heap validates a comparison: {\tt shmem\_wait\_until}.
The first step is used to measure the acknowledgment time ({\tt rt1}) between the root process and each process {\tt task}. The algorithm is using a "large" number of repetitions
% 'M`.
(M).
In the algorithm described in \cite{SK99}, this first step is made using an exchange of two-sided send and receive communications. This step is used to measure the time taken by an acknowledgment, later used in the algorithm.
Then a first broadcast is performed and process $task$ acknowledges it to the root of the broadcast (lines \ref{algo:SK:bcastinit:begin} to \ref{algo:SK:bcastinit:end}).
%%% Camille, is this next sentence incomplete?
%%% -> I added a reference and a closing parenthesis
Then comes the measurement step itself, performed $M$ times for each task (line \ref{algo:SK:measure:begin} to \ref{algo:SK:measure:end}).
%%%
The broadcast time is obtained by subtracting the acknowledgment measured in the first exchange step (line \ref{algo:SK:time}).

%% This algorithms can actually have two variants:
% - the root process and a process P exchange acks
% - process P acks to the root process, which waits for this ack to proceed
% The second version seems to work (I have the proof for both)

\addtocounter{algocf}{-1}
\begin{wrapfigure}[31]{r}{0.56\textwidth}
%\begin{figure}[htt!]
\vspace*{-.35in}    
\begin{algorithm*}[H]
\small
\resumenumbering
  \Pro{\Fwup{$task$}}{
      \tcc{Warm-up}
      \Fbcast{ $buffer, root$ }\;
      \If{ $root == rank$} {
    	\Fack{task}\;
    	\Fwforack{}\;
      } \ElseIf { $task == rank$ } {
    	\Fwforack{}\;
    	\Fack{root}\;
      }
   \KwRet\;
}
  \For{ $task \gets 0$ \KwTo $size$ \KwBy $1$ }{
      \tcc{Initialize}
      $rt1 \gets$ \FInit{ $task$ }\;
    \Fwup{ $task$ }\;
      \tcc{Measure broadcast}
      $t1 \gets $ \Fwtime{}\;  
      \For{ $i \gets 0$ \KwTo $M$ \KwBy $1$}{ \label{algo:SK:measure:begin}
      \Fbcast{ $buffer, root$ }\;
    	  \If{ $root == rank$} {
    	\Fwforack{}\;
    	  } \ElseIf { $task == rank$ } {
    	\Fack{root}\;
    	  }
      } \label{algo:SK:measure:end}
      $t2 \gets $ \Fwtime{}\;  
      \If{ $rank == task$ }{ 
      $rt2 \gets t2 - t1$\;
    	  $myt \gets rt2 - rt1$\; 
    	  $btime \gets $\Fmax{$btime, myt$}\;\label{algo:SK:time}
      }
  }
  \KwRet $btime$\;
\caption{(Part 2) One-sided communications broadcast measurement.\label{algo:SK}}
\end{algorithm*}
%\end{figure}
\end{wrapfigure}
We are performing the acknowledgment by incrementing a remote counter using an atomic
% fetch-and-inc
{\tt fetch\_and\_inc} operation and waiting for the value of this counter.
% This operation is described by algorithm \ref{algo:exchangeack}.
We are using a {\tt fetch\_and\_add} operation instead of an {\tt add} operation because, as specified by the OpenSHMEM standard, non-fetching calls may return before the operation executes on the remote process. We can wait for remote completion with a {\tt shmem\_quiet}, but we decided to use the simplest operation available; besides, although this operation goes back-and-forth between the source and the target, we are measuring this exchange in the initialization of the measurement routine.

\ignore{
\begin{algorithm}[H]
  \If{ $root == rank$ }{
	  $t \gets \mathrm{shmem\_int\_atomic\_fetch\_inc}( ack, task )$\;
	  $\mathrm{shmem\_int\_wait\_until}( ack, SHMEM\_CMP\_EQ, i )$\;
	  $*ack \gets 0$\;
  } \ElseIf{ $task == rank$ }{
    $\mathrm{shmem\_int\_wait\_until}( ack, SHMEM\_CMP\_EQ, 1 )$\;
	$*ack \gets 0$\;
	$t \gets \mathrm{shmem\_int\_atomic\_fetch\_inc}( ack, root )$\;
  }
  \caption{Exchange acknowledgments.\label{algo:exchangeack}}
\end{algorithm}
}

%% The algorithm above is only necessary of the collective operations are not ordered. They are supposed to be.

\begin{lemma} even if a remote write operation interleaves between the operations\\
{\tt shmem\_int\_wait\_until( ack,\\ SHMEM\_CMP\_EQ, 1 )}
and \verb|*ack = 0|, there cannot be two consecutive remote increment of the \verb|*ack| variable. In other words, \verb|*ack| cannot take any other value than 0 and 1 and acknowledgments sent as atomic increments are consumed by {\tt shmem\_int\_wait\_until} before another acknowledgment arrives.
\end{lemma}

\begin{proof}
If we denote:
\begin{itemize}
\item $\prec$ the \emph{happens before} relation between two events, with $a \prec b$ meaning that $a$ happens before $b$;
\item $broadcast_{ay}$: the local operation on process $y$ for the $a$th broadcast (beginning of the operation);
\item $fetch\_inc_{ay}$: the local operation on process $y$ for the $a$th {\tt shmem\_int\_atomic\_fetch\_inc} (on the source process $y$, beginning of the operation);
\item $inc_{ay}$: the increment of {\tt *ack} on process $y$ for the $a$th {\tt shmem\_int\_atomic\_fetch\_inc} (on the target process $y$);
\item $wait_{ay}$: the $a$th time process $y$ waits for the value of {\tt *ack} to be modified with {\tt shmem\_int\_wait\_until( ack, SHMEM\_CMP\_EQ, 1 )};
\item $assignment_{ay}$: the $a$th time process $y$ assigns {\tt *ack} to the value 0, hence consuming previously received acknowledgments.
\end{itemize}
%%% Camille, please look at this
%%% -> done
These operations are represented Figure \ref{fig:measure:bcast:timestamps}. Even if {\tt *ack} is incremented remotely which the broadcast operation is still in progress on the root process, it will not be read before the end of the root's participation to the broadcast. Hence, the root cannot begin the next broadcast before it is done with the current one. We need to note the fact that other processes might still be in the broadcast.
%%%

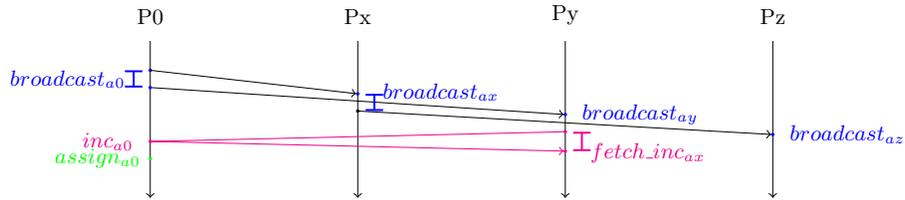
\begin{figure}
\vspace*{-.2in}
\centering
\resizebox{\linewidth}{!}{\input{figures/measure_bcast}}
\vspace*{-.1in}
\caption{Broadcast measurement}
\vspace*{-.2in}
\label{fig:measure:bcast:timestamps}
\end{figure}

The next broadcast starts after {\tt *ack} has been set back to 0. We have:
\begin{itemize}
\item on the root process:\\
$broadcast_{1root} \prec inc_{1root} \prec wait_{1root} \prec assignment_{1root} \prec broadcast_{2root} \prec inc_{2root} \prec wait_{2root} \prec assignment_{2root}$
\item on any Px\\
%$broadcast_{1x} \prec wait_{1x} \prec assignment_{1x} \prec fetch\_inc_{1x} \prec broadcast_{2x} \prec wait_{2x} \prec assignment_{2x} \prec fetch\_inc_{2x}$
$broadcast_{1x}\prec fetch\_inc_{1x} \prec broadcast_{2x} \prec fetch\_inc_{2x}$
\end{itemize}

\noindent
We also know that:

\begin{itemize}
%\item $inc_{1root} \prec assignment_{1x}$
\item $fetch\_inc_{1x} \prec inc_{1root}$
\item $inc_{1root} \prec assignment_{1root}$
\end{itemize}

% And
The OpenSHMEM standard states that: "\emph{When calling multiple subsequent collective operations on a team, the collective operations—along with any relevant team based resources—are matched across the PEs in the team based on ordering of collective routine calls}". Hence, for any two processes $x$ and $y$, $broadcast_{1x} \prec broadcast_{2y}$, so $fetch\_inc_{1x} \prec fetch\_inc_{2y}$.

Therefore, by transitivity of the $\prec$ relation:\\
$fetch\_inc_{1x} \prec assignment_{1root} \prec inc_{2root} \prec assignment_{2root}$

we can conclude that there cannot be two consecutive {\tt shmem\_int\_atomic\_fetch\_inc} on the root process with no re-assignment to 0 between them, and on any process, the acknowledgment cannot be sent while the previous or the next broadcast is in progress on other processes. $\qed$

\end{proof}

The corollary of the lemma is that acknowledgment exchanges cannot be interleaved with broadcasts (only local operations can be), and therefore 1) there is no deadlock and 2) consecutive broadcasts cannot interleave. 

%% TODO
% Need to clarify this absence of pipeline (they can but we don't care)
% Introduce wait

\ignore{

Even though the `shmem_int_atomic_fetch_inc` and `ack = 0` are two distinct operations and not a single, atomic one, the next `shmem_int_atomic_fetch_inc` cannot interleave with the `ack = 0` assignment. This is the normal order of operations:

| root                 || Px               ||
|:----------:|:-------:|:---------:|:-------:|
| operation | memory | operation | memory |
|  | `*ack = 0`  |  |`*ack = 0` |
| broadcast |`*ack = 0`  | broadcast |`*ack = 0`  |
| shmem_int_atomic_fetch_inc | `*ack = 0`  |wait until *ack == 1  |`*ack = 1` |
| wait until *ack == 1 |`*ack = 0`| *ack = 0 | `*ack = 0`   |
| wait until *ack == 1|`*ack = 1`| shmem_int_atomic_fetch_inc | `*ack = 0` |
| *ack = 0 | `*ack = 0`  | | `*ack = 0`   |
| broadcast |`*ack = 0`  | broadcast |`*ack = 0`  |
| shmem_int_atomic_fetch_inc | `*ack = 0`  |wait until *ack == 1  |`*ack = 1` |
| wait until *ack == 1 |`*ack = 0`| *ack = 0 | `*ack = 0`   |
| wait until *ack == 1|`*ack = 1`| shmem_int_atomic_fetch_inc | `*ack = 0` |
| *ack = 0 | `*ack = 0`  | | `*ack = 0`   |

If the second broadcast interleaves with the root process's assignment, we have:

| root                 || Px               ||
|:----------:|:-------:|:---------:|:-------:|
| operation | memory | operation | memory |
| wait until *ack == 1|`*ack = 1`| shmem_int_atomic_fetch_inc | `*ack = 0` |
| broadcast | `*ack = 1`  | | `*ack = 0`   |
|  *ack = 0 |`*ack = 0`  | broadcast |`*ack = 0`  |
| shmem_int_atomic_fetch_inc | `*ack = 0`  |wait until *ack == 1  |`*ack = 1` |
| wait until *ack == 1 |`*ack = 0`| *ack = 0 | `*ack = 0`   |

Hence, Px's second `shmem_int_atomic_fetch_inc` on `root` cannot happen before `root`'s first `*ack = 0` asignment.

}

%% file: figures/bcast_barrier.tex
\begin{tikzpicture}

\node (sb) at ( 0, 0 )[label={[label distance=-1mm]above:root}]{};
\node (se) at ( 0, -3 ){};
\node (tb) at ( 3, 0 )[label={[label distance=-1mm]above:Px}]{};
\node (te) at ( 3, -3){};
\node (ub) at ( 6, 0 )[label={[label distance=-1mm]above:Py}]{};
\node (ue) at ( 6, -3){};
\node (vb) at ( 9, 0 )[label={[label distance=-1mm]above:Pz}]{};
\node (ve) at ( 9, -3){};

\draw[->] (sb) -- (se);
\draw[->] (tb) -- (te);
\draw[->] (ub) -- (ue);
\draw[->] (vb) -- (ve);

\node(A) [below=2mm of sb]{};
\node(B) at ( tb |- A) {};
\node(C) at ( ub |- A) {};
\node(Cv) at ( vb |- A) {};
\node(D) [below = 3 mm of Cv] {};

\draw[color=black!60!green,dashed](A.west) rectangle  (D.east) node [midway] {barrier};

\node (E) at ( A |- D) {};
\node(F) [below = 0 mm of E,circle,inner sep=0pt,draw, fill] {};

\node (G) at ( F -| tb) {};
\node (H) at ( F -| ub) {};
\node(I) [below = 2 mm of G,circle,inner sep=0pt,draw, fill] {};
\node(J) [below = 5 mm of H,circle,inner sep=0pt,draw, fill] {};
\node(K) [below = 2 mm of F,circle,inner sep=0pt,draw, fill] {};

\draw[->] (F) -- (I);
\draw[->] (K) -- (J);

\node(N) [below = 2 mm of I,circle,inner sep=0pt,draw, fill] {};
\node (L) at ( N -| vb) {};
\node (M) [below = 2 mm of L,circle,inner sep=0pt,draw, fill] {};

\draw[->] (N) -- (M);

\node (O) [below = 1 mm of M] {};
\node (P) [below = .5 mm of O] {};
\node (Q) at ( P -| sb) {};
\node (R) [below = 1 mm of K] {};
\node (S) [below = 1 mm of N] {};

\node (T) [below = 2 mm of J] {};

\path [color=black!60!green,dashed, draw](O.east) -- (P.east) -- (Q.west) node [above,midway] {barrier} -- (R.west) -- (S.center) -- (T.center) -- (O.center);

\node(beginb) [left=2mm of A]{};
\node(endb) at (beginb |- D){};

\draw [|-|, color=blue,decorate,decoration={snake,amplitude=.3mm,segment length=2mm,post length=1mm}](endb.center) -- (beginb.center)
node [midway,above,sloped] {m1};

\node(endt) at (beginb |- Q){};
\node(begint) at (beginb |- F){};

\draw [|-|, color=violet,decorate,decoration={snake,amplitude=.3mm,segment length=2mm,post length=1mm}](endt.center) -- (begint.center)
node [midway,above,sloped] {m2};

\node(U)  [right=2mm of O]{};
\node(endq) at (U |- P){};

\draw [|-|, color=magenta,decorate,decoration={snake,amplitude=.3mm,segment length=2mm,post length=1mm}](U.center) -- (endq.center)
node [midway,above,sloped] {m3};

\end{tikzpicture}

%% file: figures/measure_bcast.tex
\begin{tikzpicture}

\node (sb) at ( 0, 0 )[label={[label distance=-1mm]above:P0}]{};
\node (se) at ( 0, -2.5 ){};
\node (tb) at ( 3, 0 )[label={[label distance=-1mm]above:Px}]{};
\node (te) at ( 3, -2.5){};
\node (ub) at ( 6, 0 )[label={[label distance=-1mm]above:Py}]{};
\node (ue) at ( 6, -2.5){};
\node (vb) at ( 9, 0 )[label={[label distance=-1mm]above:Pz}]{};
\node (ve) at ( 9, -2.5){};

\draw[->] (sb) -- (se);
\draw[->] (tb) -- (te);
\draw[->] (ub) -- (ue);
\draw[->] (vb) -- (ve);

\node(A) [below=4mm of sb,circle,inner sep=0pt,draw, fill, color=blue]{};
\node(B) at ( tb |- A) {};
\node(C) at ( ub |- A) {};
\node(D) at ( vb |- A) {};

\node(F) [below = 2 mm of A,circle,inner sep=0pt,draw, fill, color=blue] {};

\node (G) at ( F -| tb) {};
\node (H) at ( F -| ub) {};

\node(I) [below = 2 mm of B,circle,inner sep=0pt,draw, fill, color=blue] {};
\node(J) [below = 5 mm of C,circle,inner sep=0pt,draw, fill,color=blue] {};

%\node(K) [below = 2 mm of F,circle,inner sep=0pt,draw, fill] {};

\draw[->] (A.center) -- (I);
\draw[->] (F) -- (J);

\node(N) [below = 2 mm of I,circle,inner sep=0pt,draw, fill] {};
\node (L) at ( N -| vb) {};
\node (M) [below = 2 mm of L,circle,inner sep=0pt,draw, fill,color=blue] {};

\draw[->] (N) -- (M);

\node (begin0)[left=1mm of A] {};
\node (end0)[left=1mm of F] {};
\draw [|-|, color=blue,thick](end0.center) -- (begin0.center) node [midway,left] {$broadcast_{a0}$};

\node (beginx)[right=1mm of I] {};
\node (endx)[right=1mm of N] {};
\draw [|-|, color=blue,thick](endx.center) -- (beginx.center) node [pos=1.1,right] {$broadcast_{ax}$};

\node [color=blue,thick,right=1mm of J]  {$broadcast_{ay}$};
\node [color=blue,thick,right=1mm of M]  {$broadcast_{az}$};

\node (O) [below = 2 mm of J,circle,inner sep=0pt,draw, fill,color=magenta] {};
\node (P) at ( O -| sb) {};
\node (Q) [below =0 mm of P,circle,inner sep=0pt,draw, fill,color=magenta] {};
\node (R) at ( Q -| ub) {};
\node (S) [below =0 mm of R,circle,inner sep=0pt,draw, fill,color=magenta] {};

\draw[->,color=magenta] (O) -- (Q) -- (S);

\node (beginxx)[right=1mm of O] {};
\node (endxx)[right=1mm of S] {};
\draw [|-|, color=magenta,thick](endxx.center) -- (beginxx.center) node [pos=-.1,right] {$fetch\_inc_{ax}$};

\node [color=magenta,thick,left=1mm of Q]  {$inc_{a0}$};

\node (Q) [below =2 mm of Q,circle,inner sep=0pt,draw, fill,color=green,label=left:\color{green}{$assign_{a0}$}] {};

\end{tikzpicture}

%% file: measureothers.tex
%\subsection{Other OpenSHMEM routines}
%\label{sec:measure:others}

\subsection{Fine-grain measurements}
\label{sec:measure:fine}

There are important concerns we needed to pay
attention to when updating \skampi for making fine-grain measurements.

\begin{minipage}{0.48\textwidth}
\begin{algorithm}[H]\small
    $t1 \gets \mathrm{wtime}()$\;
   \For{$i\gets0$ \KwTo $iterations$ \KwBy $1$}{
    $\mathrm{shmem\_putmem}( ... )$\;
    $\mathrm{shmem\_quiet}()$\;
   }
    $ttime \gets \mathrm{wtime}() - t1$\;
  \Return $ttime / iterations$\; 
      \DontPrintSemicolon
            \nonl \;
      \PrintSemicolon
  \caption{Timing outside.\label{algo:timing2}}
\end{algorithm}
\end{minipage}
\begin{minipage}{0.48\textwidth}
\begin{algorithm}[H]\small
   $ttime \gets 0$\;
   \For{$i\gets0$ \KwTo $iterations$ \KwBy $1$}{
    $t1 \gets \mathrm{wtime}()$\;
    $\mathrm{shmem\_putmem}( ... )$\;
    $\mathrm{shmem\_quiet}()$\;
    $ttime \gets ttime + \mathrm{wtime}() - t1$\;
  }
  \Return $ttime / iterations$\; 
  \caption{Timing inside.\label{algo:timing1}}
\end{algorithm}
\end{minipage}

\paragraph{Measurement disturbance.} The timing function in \skampi
(a call to PAPI's timing routine) takes a time of the same order of magnitude or, in some cases, higher that the time taken by some of the functions we are measuring.
Hence, we want to minimize the number of calls to the timing function
during a measurement.
We observed very significant differences between the times obtained using Algorithm \ref{algo:timing1} and Algorithm \ref{algo:timing2}. A lot of calls to the timing function, which, if using an external function such as PAPI, we might not be able to inline, is causing very significant disturbance to the measurement.

\begin{minipage}{0.48\textwidth}
\begin{algorithm}[H]\small
   $ttime \gets 0$\;
   \For{$i\gets0$ \KwTo $iterations$ \KwBy $1$}{
    $\mathrm{shmem\_putmem}( ... )$\;
    $t1 \gets \mathrm{wtime}()$\;
    $\mathrm{shmem\_quiet}()$\;
    $ttime \gets ttime + \mathrm{wtime}() - t1$\;
  }
  \Return $ttime / iterations$\; 
  \caption{Timing {\tt shmem\_quiet}.\label{algo:timing:nb1}}
\end{algorithm}
\end{minipage}
\begin{minipage}{0.48\textwidth}
\begin{algorithm}[H]\small
    $tpost = \mathrm{get\_post\_time}()$\;
    $t1 \gets \mathrm{wtime}()$\;
   \For{$i\gets0$ \KwTo $iterations$ \KwBy $1$}{
    $\mathrm{shmem\_putmem}( ... )$\;
    $\mathrm{shmem\_quiet}()$\;
   }
    $ttime \gets \mathrm{wtime}() - t1$\;
  \Return $ttime / iterations  - tpost$\; 
  \caption{Subtraction method.\label{algo:timing:nb2}}
\end{algorithm}
\end{minipage}

\paragraph{Separating calls.} Some functions we are measuring can be called only
% along
in the context of another function.
For instance, if we want to measure the time spent waiting for a non-blocking communication to complete,
% (\verb|shmem_quiet|),
we need to post a non-blocking communication before. We cannot post a set of non-blocking communications and call \verb|shmem_quiet| in a loop, because it waits for completion of all the outstanding non-blocking communications at the same time. Therefore, each \verb|shmem_quiet| must correspond to a previously posted non-blocking communication. However, we cannot isolate it on our measurement such as described by Algorithm \ref{algo:timing:nb1}, for the reasons presented in the previous paragraph. Therefore, we are initializing the measurement by measuring the time taken by the routine that posts the non-blocking communication, measuring the whole loop, and subtracting the post time from the result (Algorithm \ref{algo:timing:nb2}).

\paragraph{Stability.} The aforementioned methods rely on subtracting values measured during the initialization of the measurement. Therefore, the experimental conditions must remain stable through the measurement. For instance, we noticed instabilities on machines that were being used by multiple users at the same time, while the measurements were quite stable on nodes used in exclusive mode. Moreover, SKaMPI calls each measuring function multiple times and keeps calling them until the standard deviation between measurements is
% not  %%% Camille, I commented out this 'not' because it made more sense.
%% Good catch!!
small enough. However, we found significant improvement in the stability between experiments when each measurement function was, itself, performing a significant number of measurements and returning their mean.

\paragraph{Busy wait.} In order to avoid voluntary context switches, we are not using a \verb|sleep| to wait while the communication is progressing in the background.
Instead, we are performing a computation operation that takes
the same time, but does not involve the operating system.
% We
Our approach is to increment a variable in a loop and we avoid compiler
optimization by inserting an call to an empty assembly instruction (\verb|asm("")|) in the loop.

%% file: expe.tex
\section{Experimental evaluation}
\label{sec:expe}

We used our extended SKaMPI on the Summit supercomputer, which features 4\,608 two-socket IBM POWER9 nodes, 6 Nvidia V100 GPUs per node and 512GB of DDR4 plus 96GB of HBM2 per node. The network is a Mellanox EDR 100G InfiniBand non-blocking fat tree. We used the provided IBM Spectrum MPI and OpenSHMEM library version 10.3.1.02rtm0 and the IBM XL compiler V16.1.1. The input files used to run these experiments are available along with the SKaMPI source file.
% %%% Camille, just check that the next sentence is correct.
The remainder of this section discusses selected outcomes from the full set of SKaMPI-OpenSHMEM results.

% TODO if we have space left at the end, I can run experiments on Grid5000 and show
% not so good overlap capabilities.
\subsection{Loop measurement granularity}
\label{sec:expe:measu}

In section \ref{sec:measure:fine} we mentioned the importance of how loops are measured. We compared the time returned by the functions that measure a non-blocking put (\verb|shmem_put| immediately followed by \verb|shmem_quiet|, \verb|shmem_put| and \verb|shmem_quiet| measured separately). We can see that iteration-level measurement introduce a very significant latency, which is not visible for longer measurements that are not latency-bound (higher buffer sizes).

\begin{figure}
\vspace*{-.2in}
\centering
\begin{subfigure}[b]{0.49\textwidth}
\includegraphics[width=\linewidth]{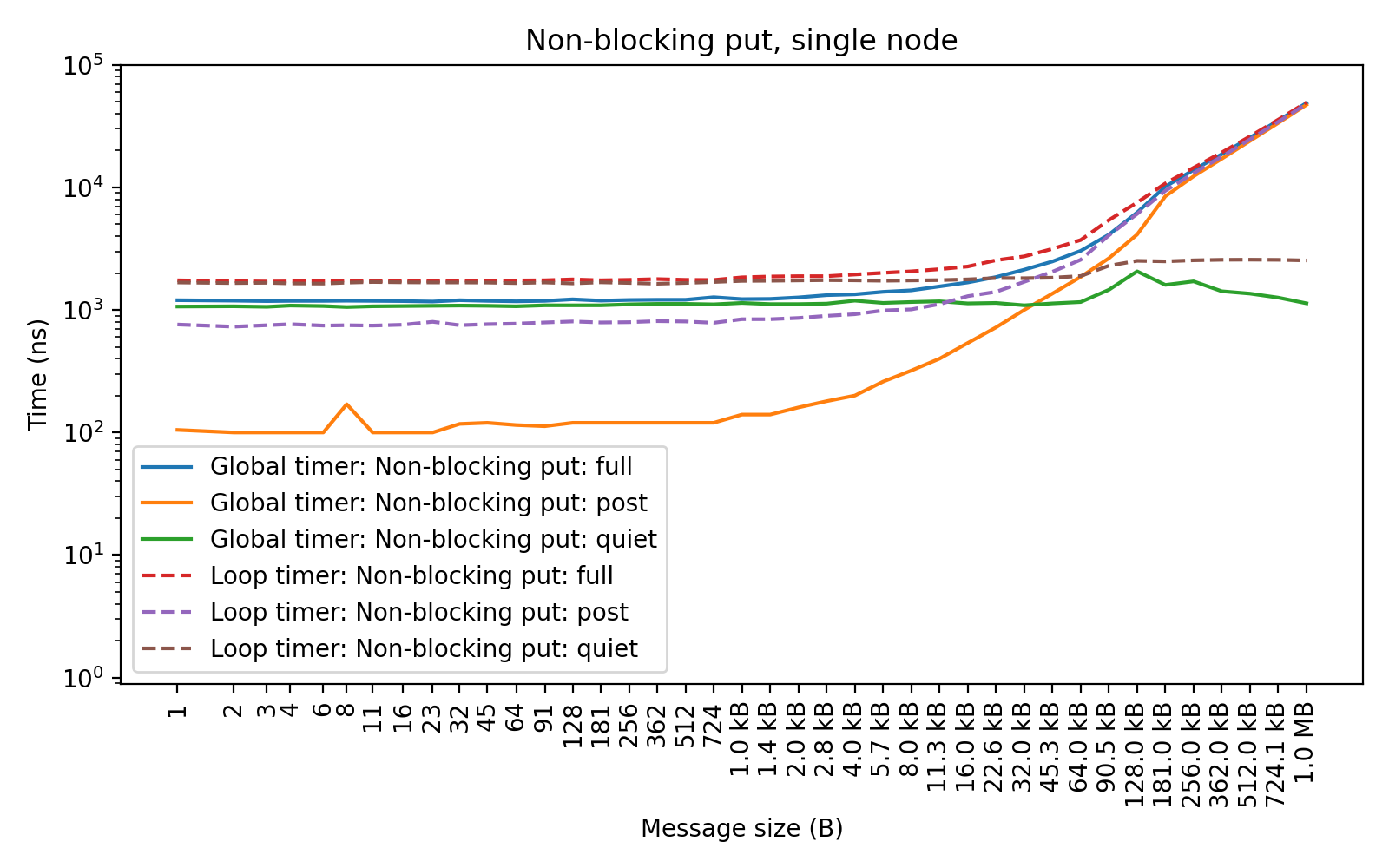}
\caption{\label{fig:exp:fine:1node}Single node.}
\end{subfigure}
\hfill
\begin{subfigure}[b]{0.49\textwidth}
\includegraphics[width=\linewidth]{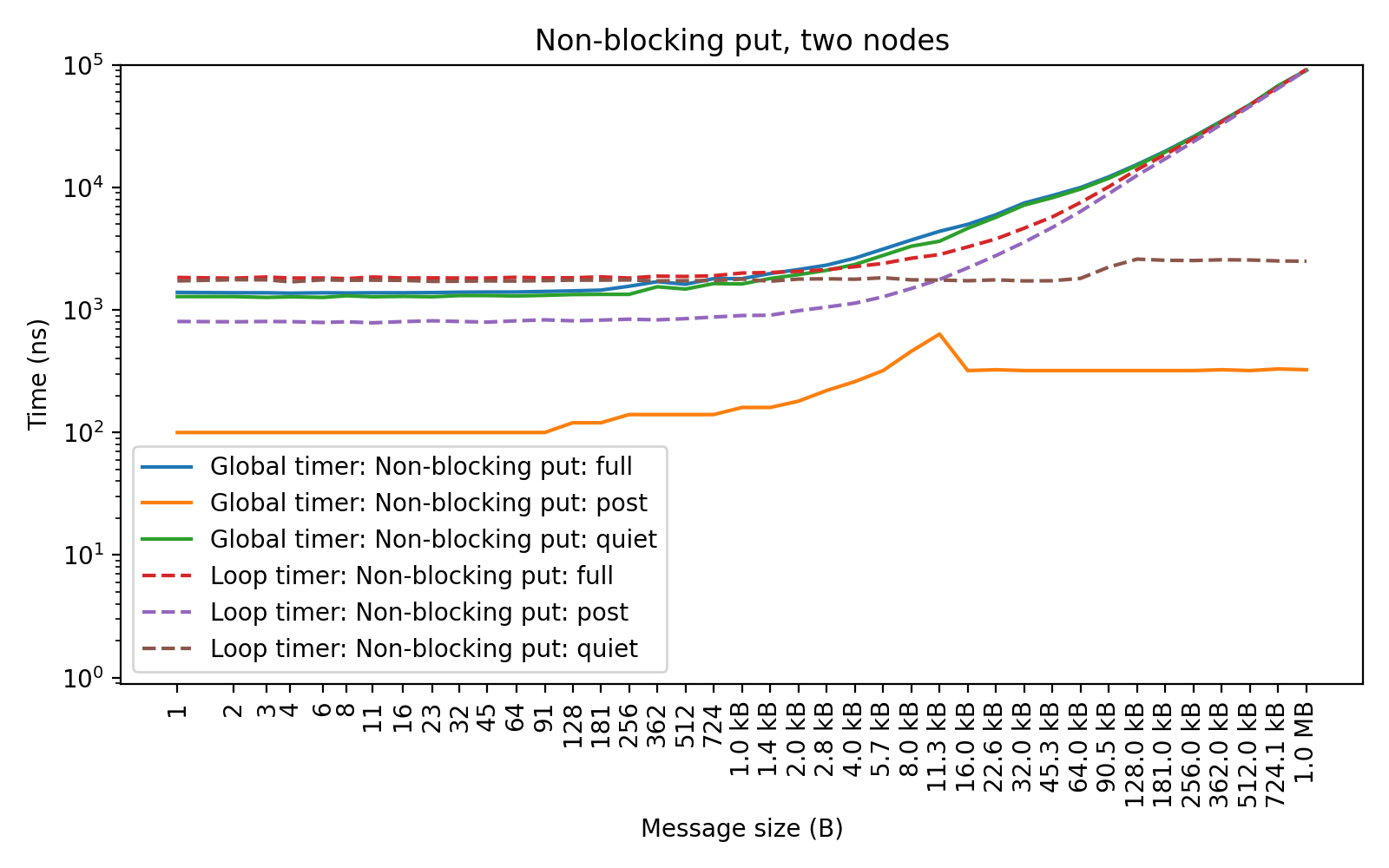}
\caption{\label{fig:exp:fine:2nodes}Two nodes.}
\end{subfigure}
\caption{\label{fig:exp:fine}Measurement granularity: iteration vs global loop timer, on the elements of a non-blocking put.}
%\vspace*{-.4in}
\end{figure}

\begin{figure}
\vspace*{-.2in}
\centering
\begin{subfigure}[b]{0.49\textwidth}
\includegraphics[width=\linewidth]{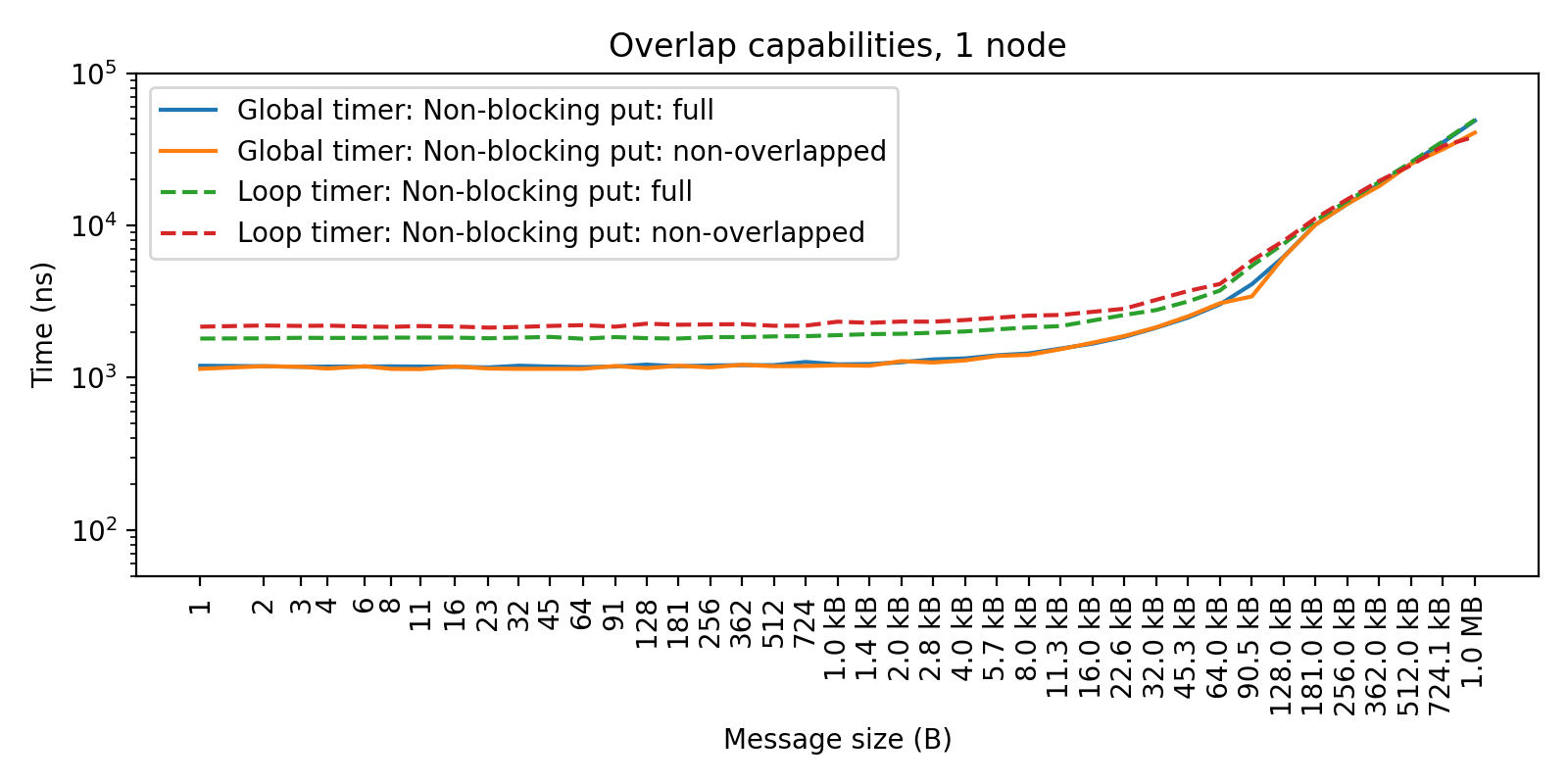}
\caption{\label{fig:exp:fine:overlap:1node}Single node.}
\end{subfigure}
\hfill
\begin{subfigure}[b]{0.49\textwidth}
\includegraphics[width=\linewidth]{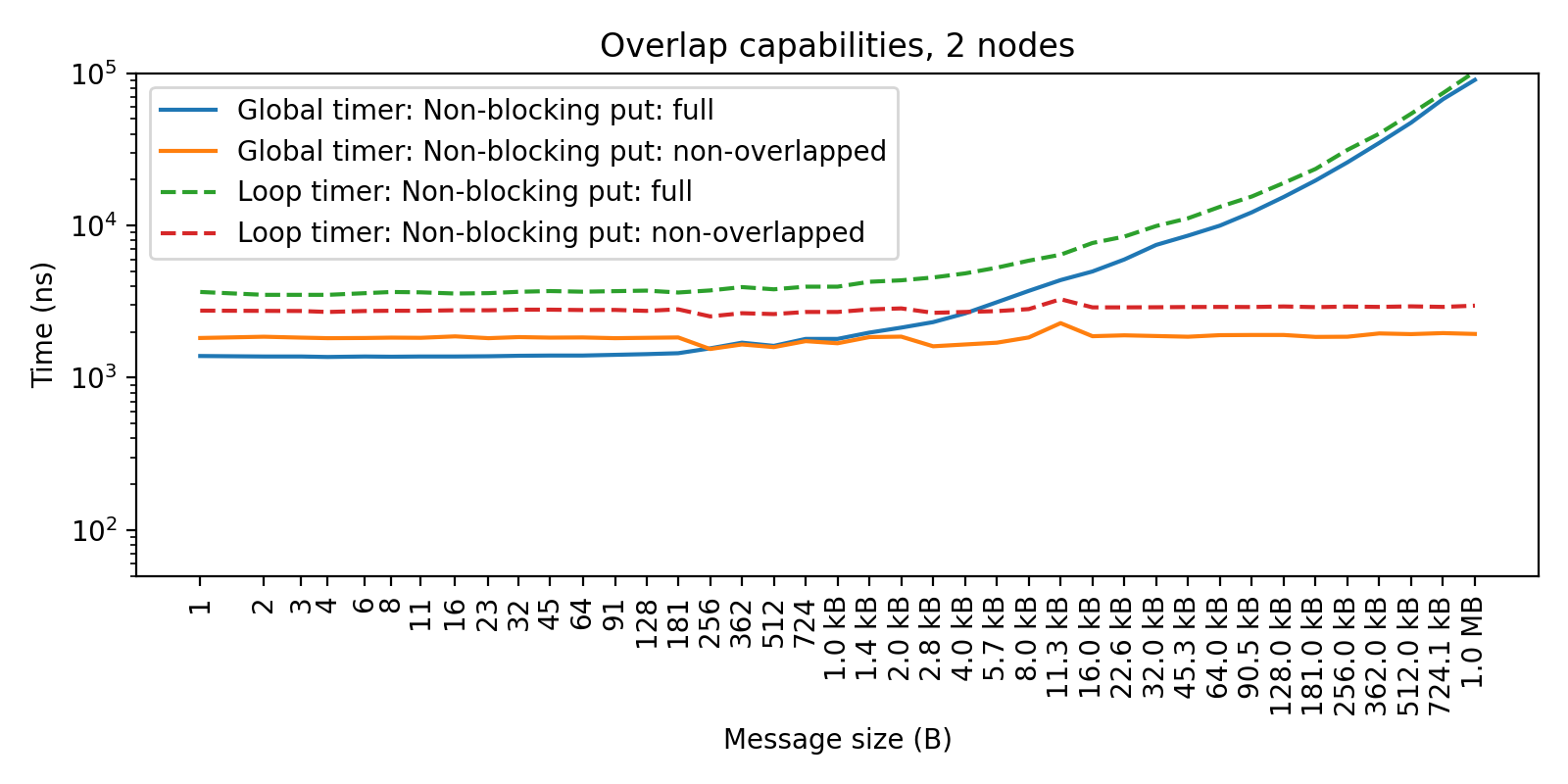}
\caption{\label{fig:exp:fine:overlap:2nodes}Two nodes.}
\end{subfigure}
\caption{\label{fig:exp:fine:overlap}Measurement granularity: iteration vs global loop timer, on the overlap capabilities of a non-blocking put.}
%\vspace*{-.4in}
\end{figure}

Similarity, we compared these time measurement strategies on the measurement of the overlap capabilities of non-blocking communications. Iteration-level measurement
% use
uses four calls to the timing routines in each iteration of the measurement loop: before and after posting the non-blocking communication, and before and after calling \verb|shmem_quiet| to wait for its completion. The global loop measurement
% measures
times the whole loop and subtracts the time assumed to be taken
by the computation used to (try to) overlap
% (or try to)
the computation. As discussed in section \ref{sec:measure:fine}, it relies on the hypothesis that this time will be stable throughout the measurement. However, as we can see Figure \ref{fig:exp:fine:overlap}, the latency introduced by the iteration-level measurement strategy is such that the numbers returned by this method are too far from reality for small messages. 

% As a consequence, of course,
% the measurements taken in the previous subsections of this paper have been measured
% using a global timer, as described in section \ref{sec:measure:fine}.
% %%% Camille, check that the following is correct.
%Consequently, we re-measured these experiments using our
Consequently, the results presented in this section were measured using our
global timer approach (as described in section \ref{sec:measure:fine}).
The results are shown in the figures.  It can be seen that the global timer measurements
produced smaller and more reliable values versus iteration timers.

\begin{figure}
\centering
\begin{subfigure}[b]{0.49\textwidth}
\includegraphics[width=\linewidth]{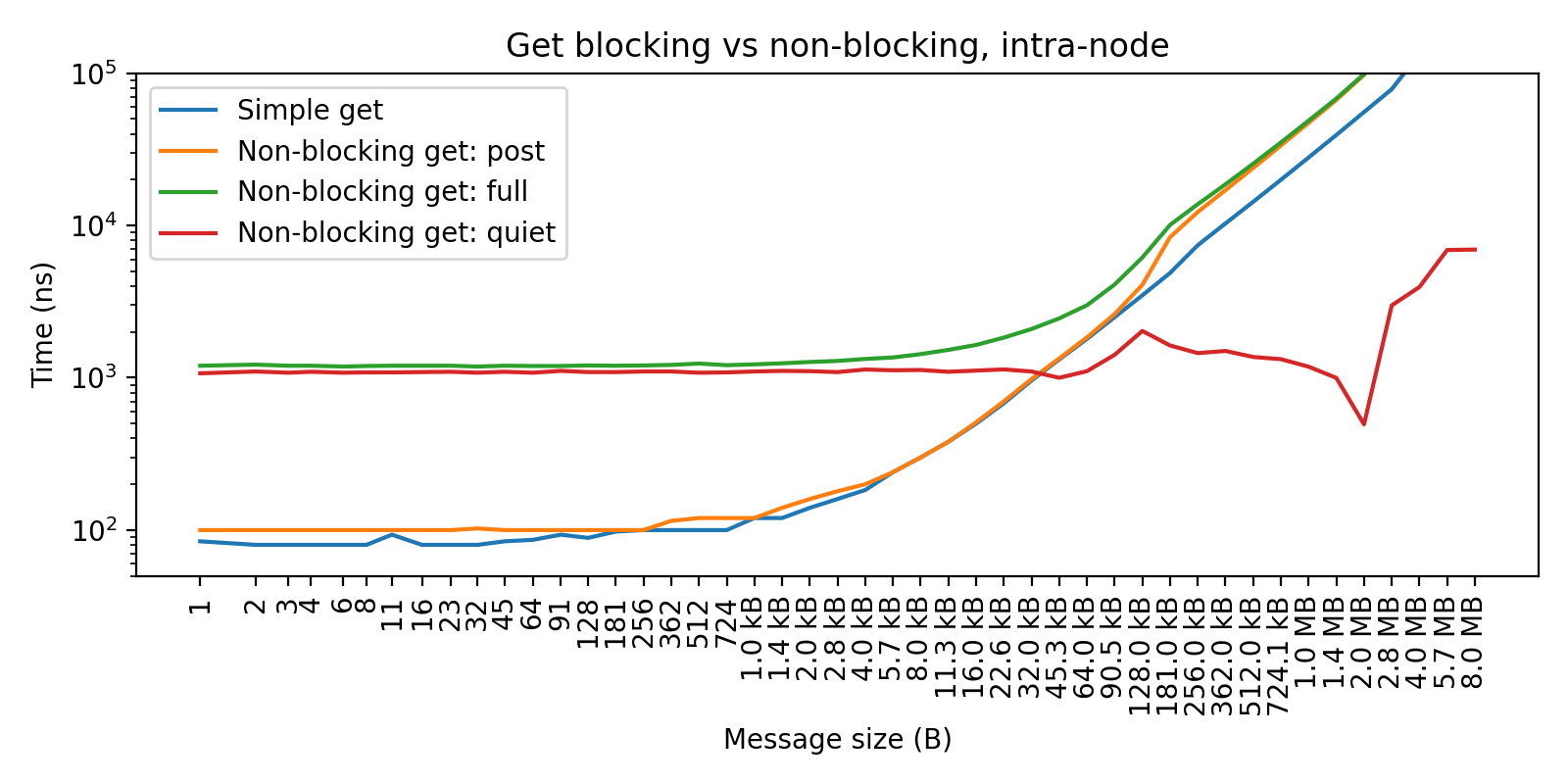}
\caption{\label{fig:exp:p2p:1node:get}Get, intra-node}
\end{subfigure}
\hfill
\begin{subfigure}[b]{0.49\textwidth}
\includegraphics[width=\linewidth]{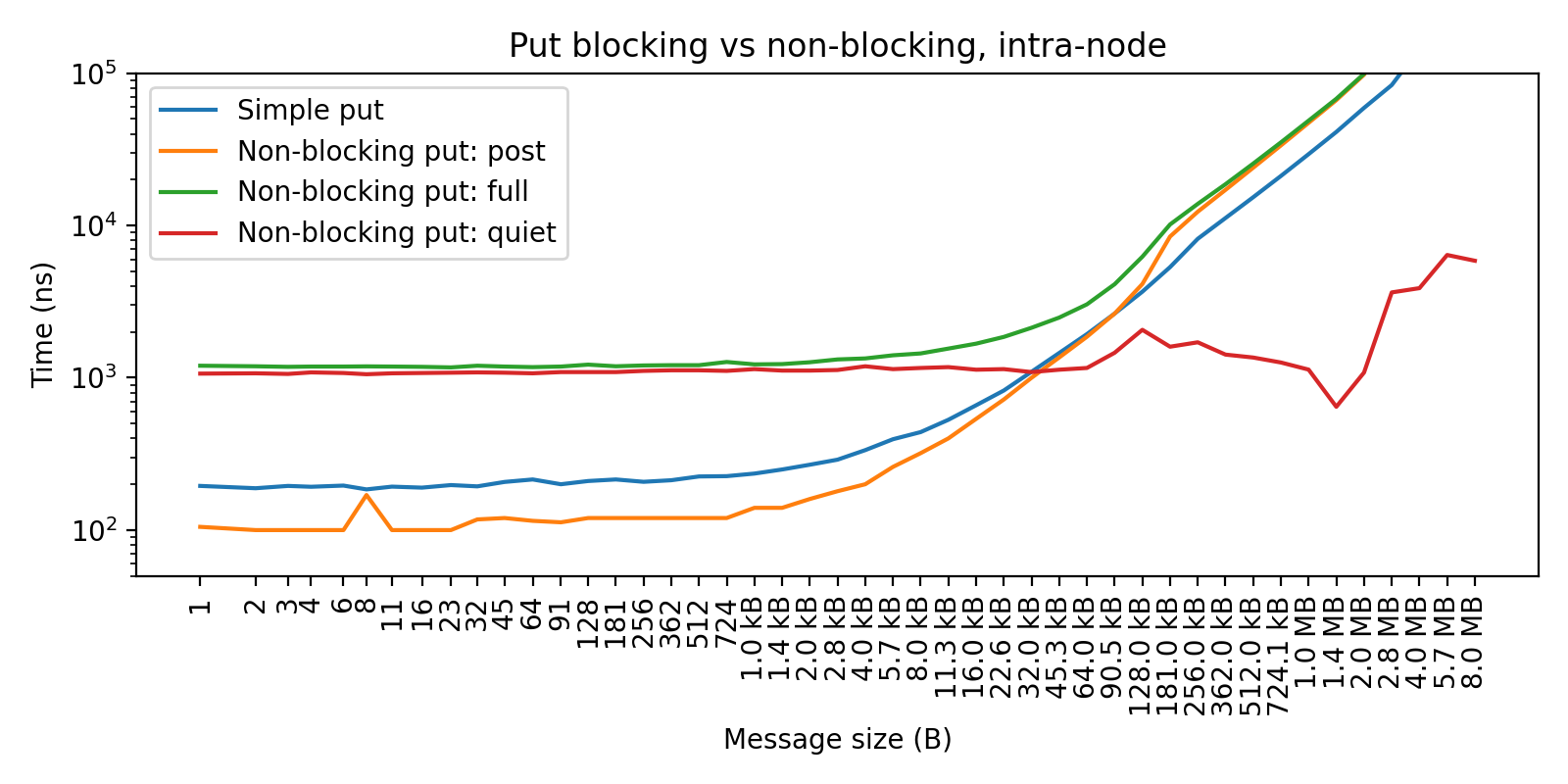}
\caption{\label{fig:exp:p2p:1node:put}Put, intra-node}
\end{subfigure}
\begin{subfigure}[b]{0.49\textwidth}
\includegraphics[width=\linewidth]{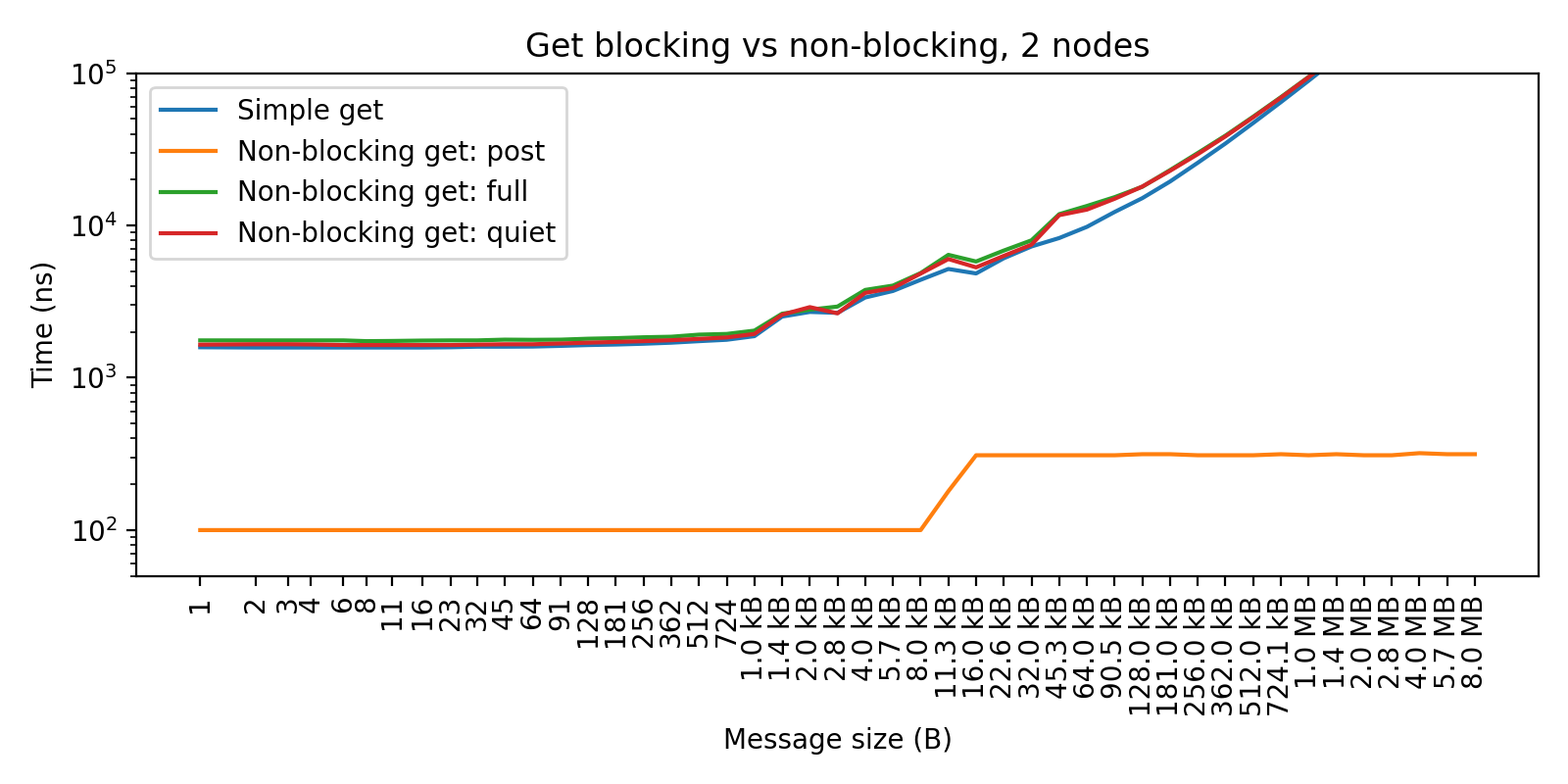}
\caption{\label{fig:exp:p2p:2nodes:get}Get, inter-node}
\end{subfigure}
\hfill
\begin{subfigure}[b]{0.49\textwidth}
\includegraphics[width=\linewidth]{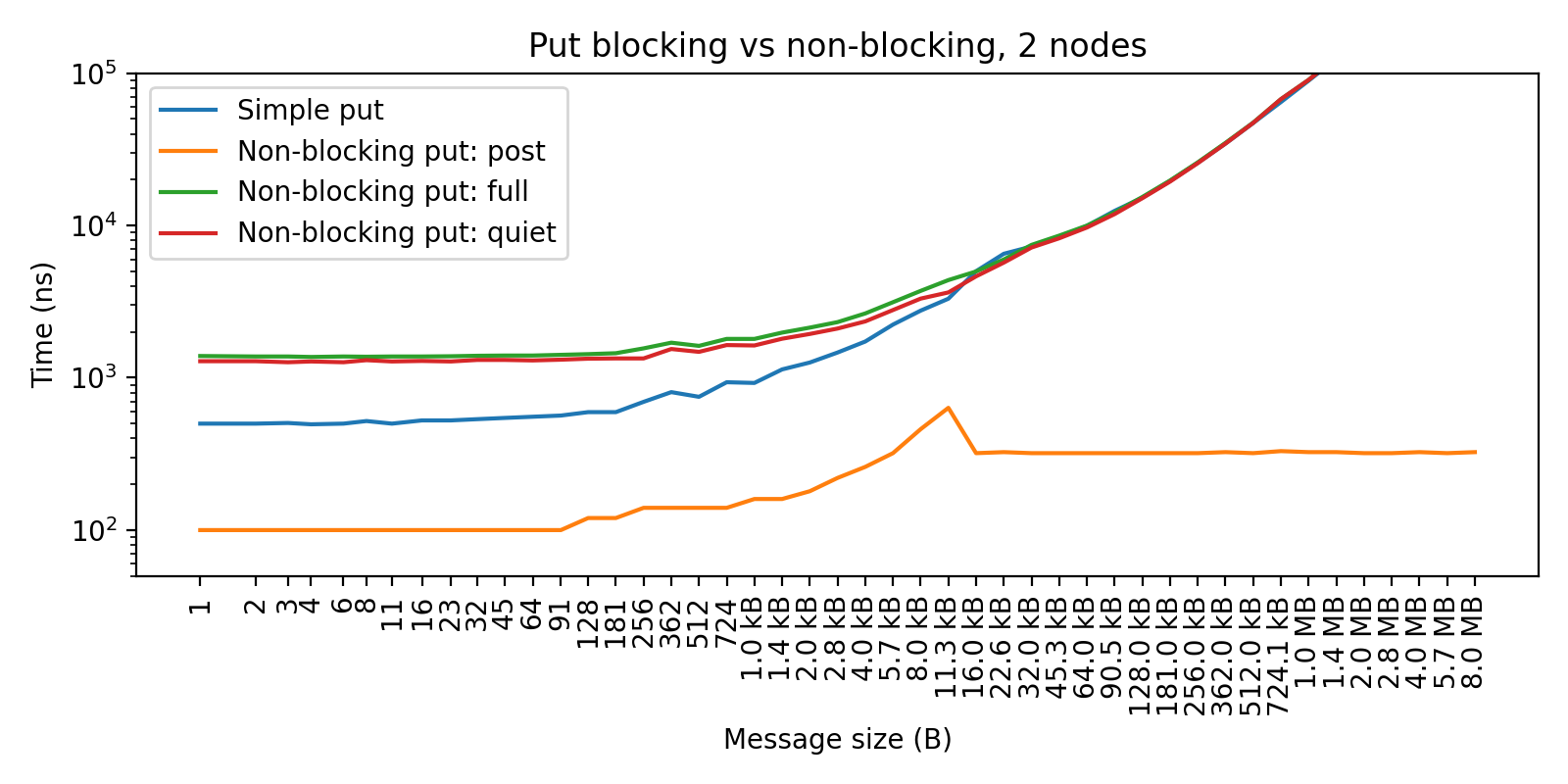}
\caption{\label{fig:exp:p2p:2nodes:put}Put, inter-node}
\end{subfigure}

\vspace*{-.1in}
\caption{\label{fig:exp:p2p}Point-to-point communication performance breakdown.}
\vspace*{-.2in}
\end{figure}

\subsection{Point-to-point communications: blocking vs non-blocking}
\label{sec:expe:p2p}

We measured the performance difference between a blocking communication and a non-blocking communication, and the time it takes to wait for completion of the communication. For instance, Figure \ref{fig:exp:p2p} shows the communication performance on a single node and between two nodes. We can see that blocking communications have a smaller latency than a non-blocking communication followed by a  {\tt shmem\_quiet} that waits for its completion. We can also see the breakdown between how much time is spent posting the non-blocking communication and how much is spent waiting for its completion.

\ignore{
Figures \ref{fig:exp:p2p:1node} and \ref{fig:exp:p2p:2nodes} show the performance we obtain between two processes running in
% respectively
a single node and on two nodes, respectively.
We measured the time to post the non-blocking communication, the time to wait for its completion with {\tt shmem\_quiet} and the time to perform the whole operation. As expected, we can see that the time to perform a simple communication ({\tt shmem\_put} or {\tt shmem\_get} followed by {\tt shmem\_quiet}) is faster
% that
than a non-blocking operation. 
}

\ignore{
\begin{figure}
\centering
\begin{subfigure}[b]{0.48\textwidth}
\includegraphics[width=\linewidth]{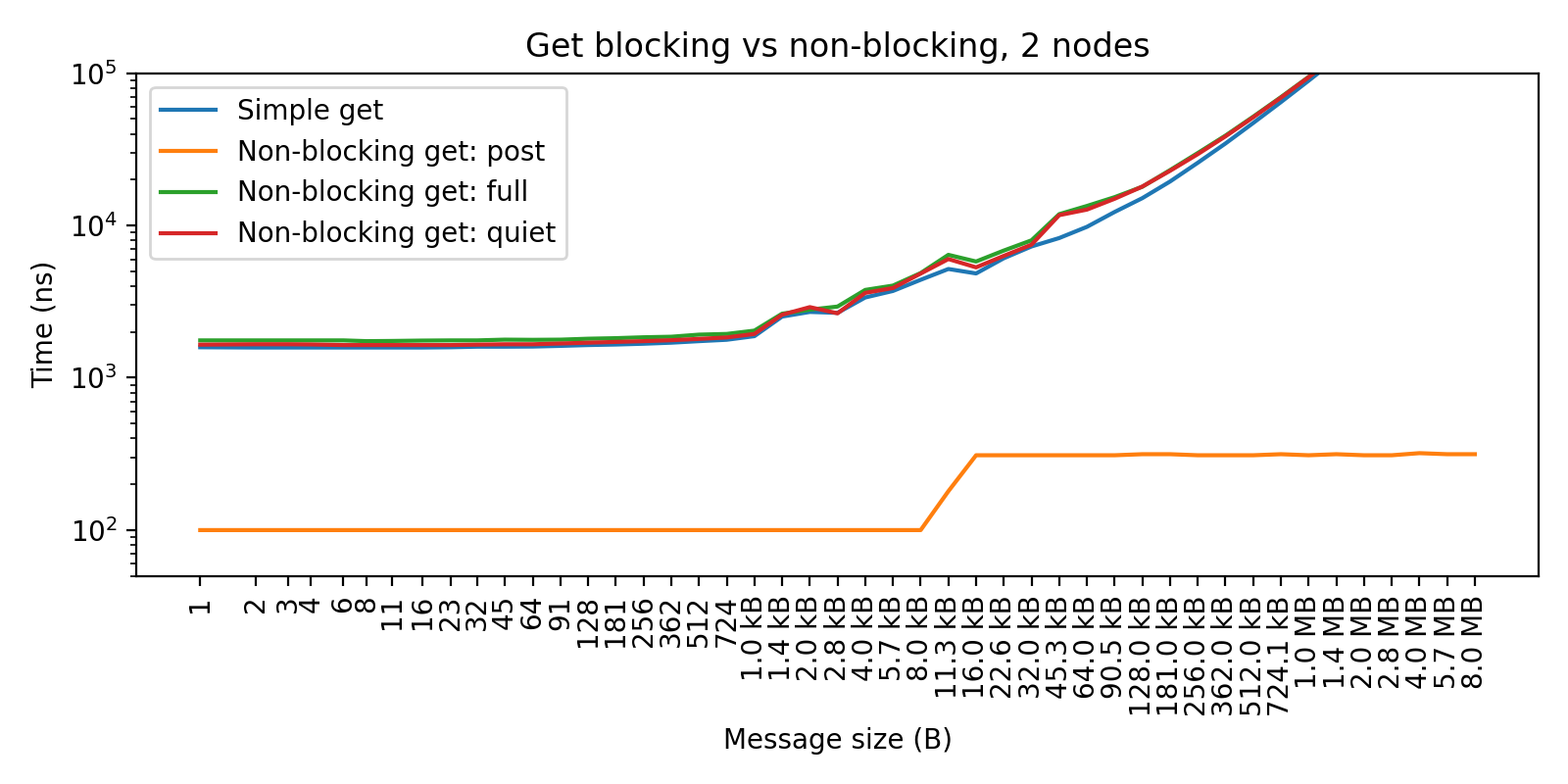}
\caption{\label{fig:exp:p2p:2nodes:get}Get}
\end{subfigure}
\hfill
\begin{subfigure}[b]{0.48\textwidth}
\includegraphics[width=\linewidth]{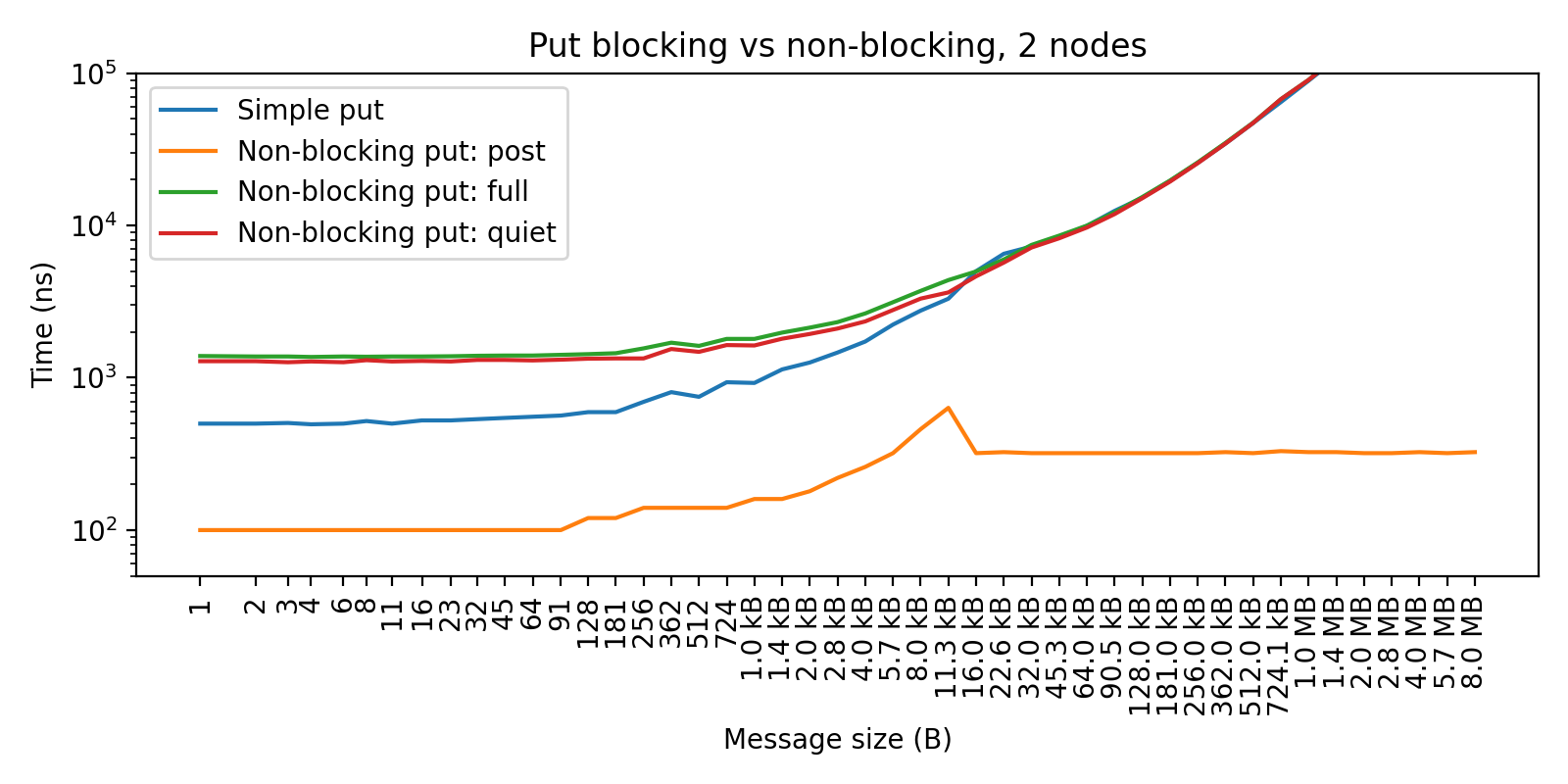}
\caption{\label{fig:exp:p2p:2nodes:put}Put}
\end{subfigure}
\caption{\label{fig:exp:p2p:2nodes}Point-to-point communication performance between two nodes.}
\end{figure}
}

\ignore{
overlap:
    /* Perform the non-blocking operations */
    for (i=0; i<iterations; i++) {
        t1 = wtime();
        shmem_putmem_nbi( sym, get_send_buffer(), count,  (rank + 1 ) % size );
        t2 = wtime();
        ttime += (t2 - t1);
        usleep( 2*btime );

        /* This is what we are measuring */
        t1 = wtime();
        shmem_quiet();
        t2 = wtime();
        ttime += (t2 - t1);
    }
    
}

\subsection{Point-to-point communications: overlap capabilities}

\begin{figure}
\centering
\begin{subfigure}[b]{0.49\textwidth}
\includegraphics[width=\linewidth]{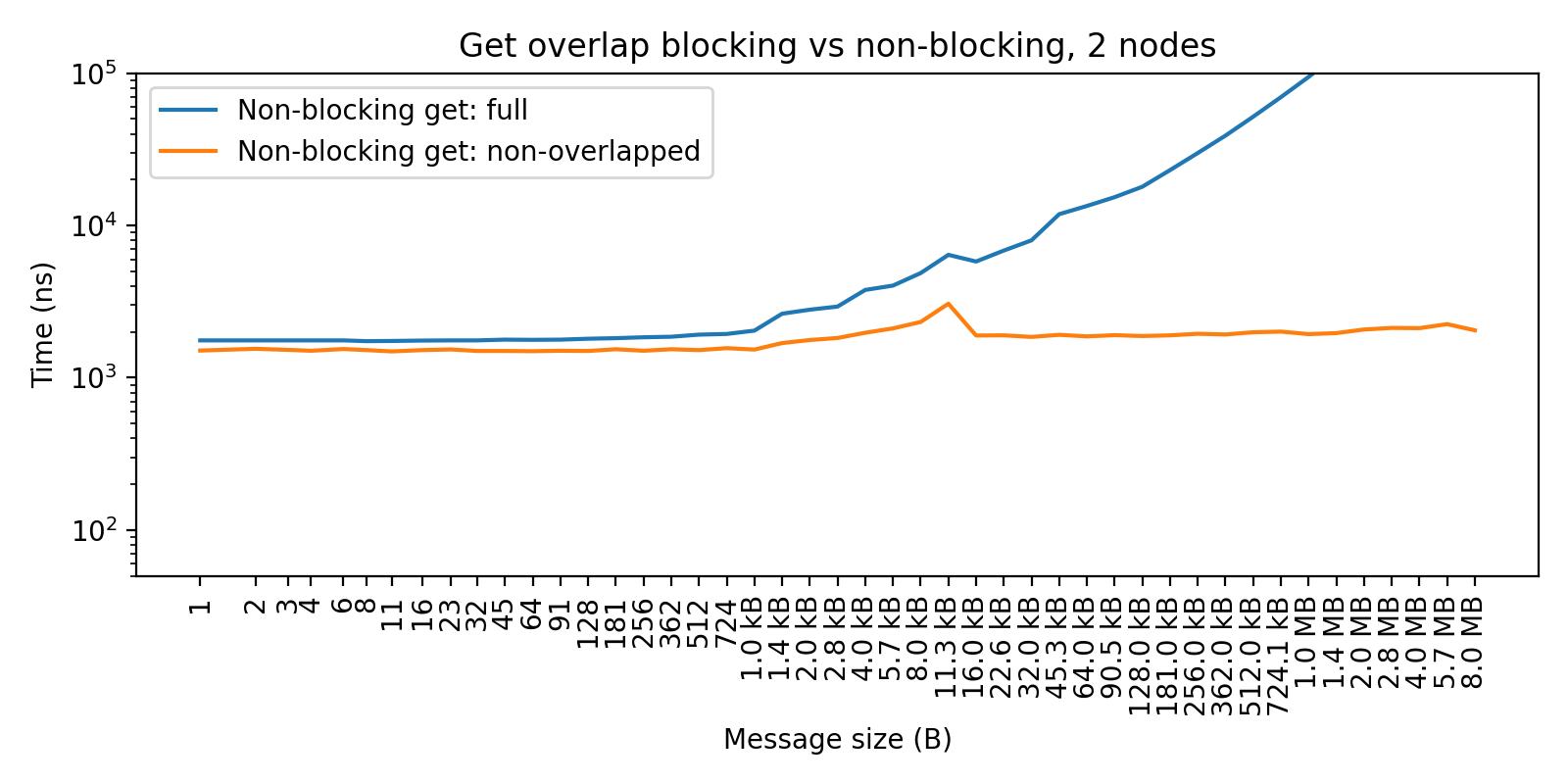}
\caption{\label{fig:exp:overlap:2nodes:get}Get, inter-nodes}
\end{subfigure}
\hfill
\begin{subfigure}[b]{0.49\textwidth}
\includegraphics[width=\linewidth]{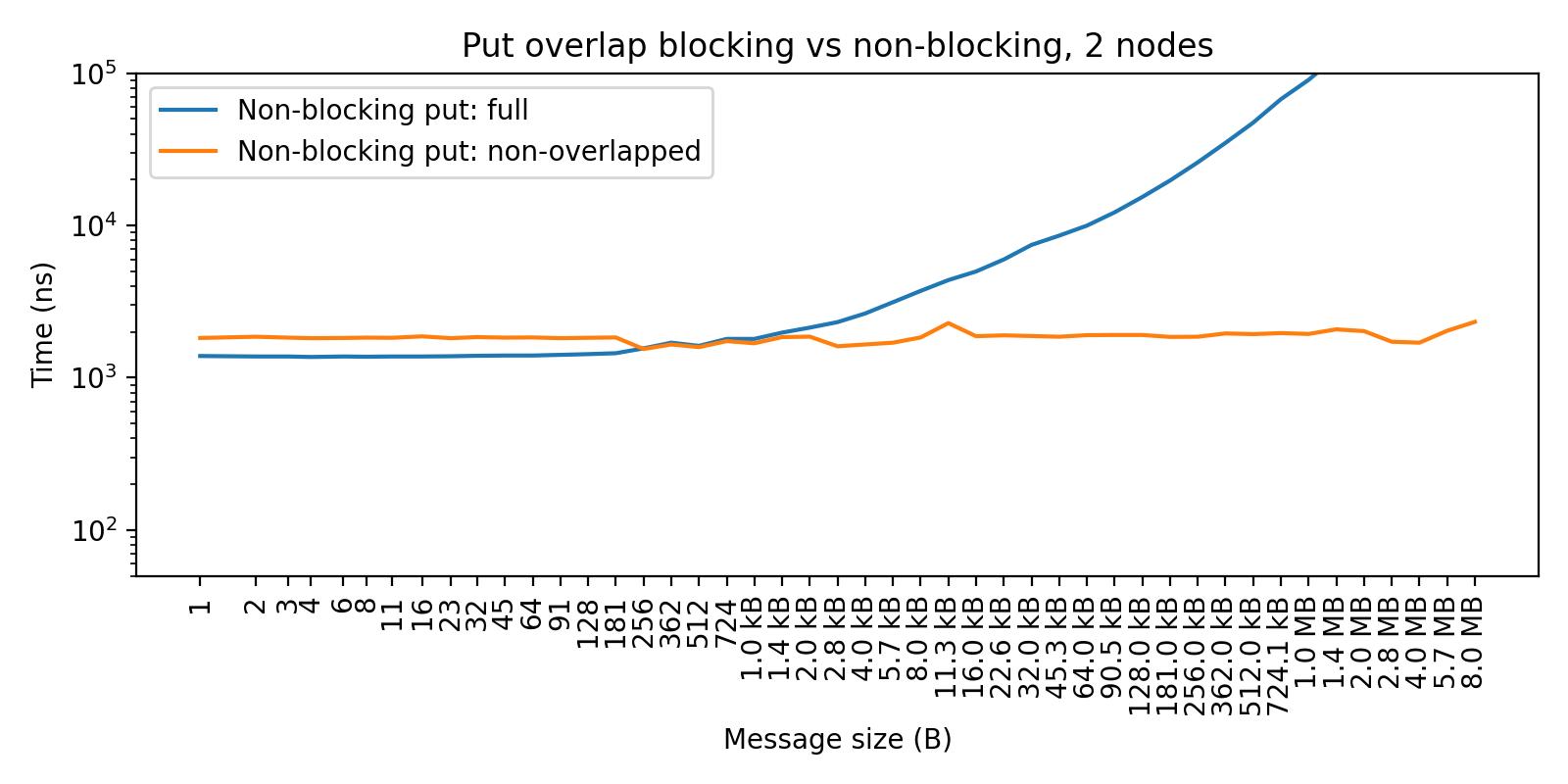}
\caption{\label{fig:exp:overlap:2nodes:put}Put, inter-nodes}
\end{subfigure}\\
\begin{subfigure}[b]{0.49\textwidth}
\includegraphics[width=\linewidth]{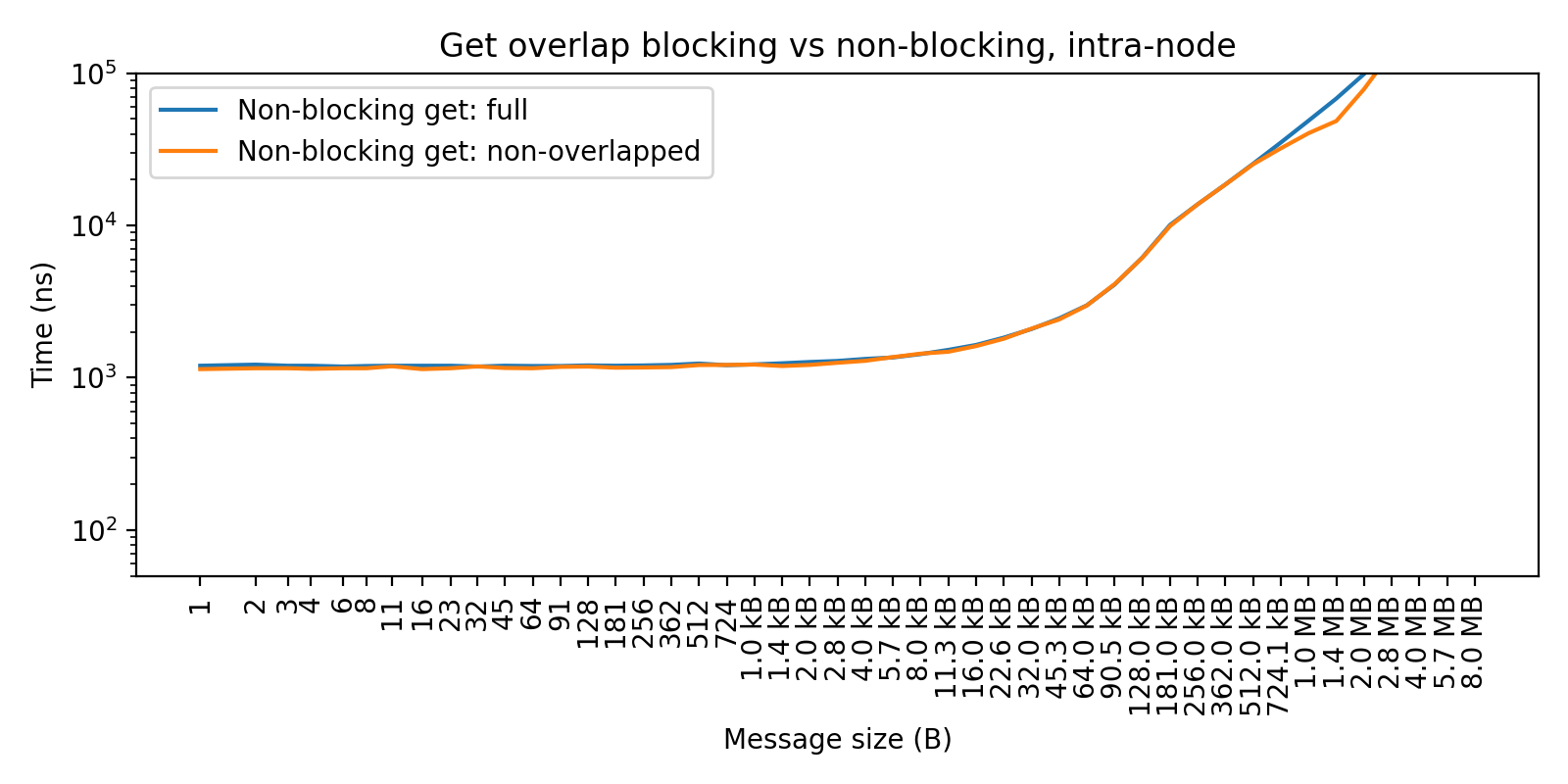}
\caption{\label{fig:exp:overlap:1node:get}Get, intra-node}
\end{subfigure}
\hfill
\begin{subfigure}[b]{0.49\textwidth}
\includegraphics[width=\linewidth]{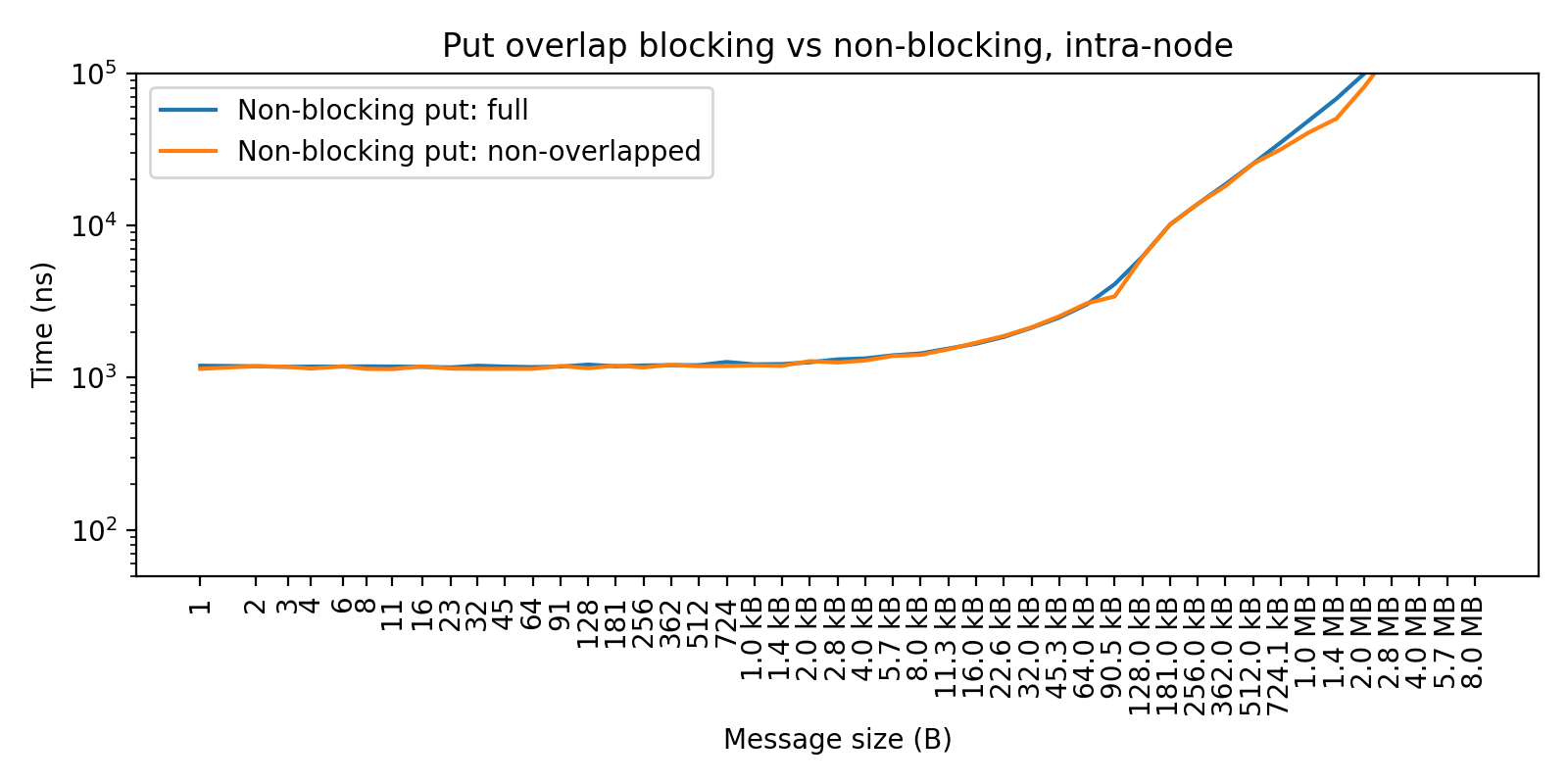}
\caption{\label{fig:exp:overlap:1node:put}Put, intra-nodes}
\end{subfigure}

\caption{\label{fig:exp:overlap}Overlap capabilities of point-to-point communications.}
\vspace*{-.2in}
\end{figure}

We can also use SKaMPI to measure the overlapping capabilities of the OpenSHMEM library, as described in section \ref{sec:measure:p2p}. Figure \ref{fig:exp:overlap} shows how much time is spent in a complete non-blocking operation ({\tt shmem\_put} or {\tt shmem\_get} immediately followed by {\tt shmem\_quiet}) and how much is spent in these operations separated by
% a
some computation. 
We can see on Figures \ref{fig:exp:overlap:2nodes:put} and \ref{fig:exp:overlap:2nodes:get} that, since the interconnection network used on the Summit machine can make the communication progress in the background, it achieves good overlap between communication and computation. The time spent in communication routines is constant (called "non-overlapped") on the figure, corresponding to the time spent in the communication routines. On the other hand, intra-node communications cannot progress in the background.
% so
% We can see
This can be seen in Figure  \ref{fig:exp:overlap:1node:put} and \ref{fig:exp:overlap:1node:get}
% that
where the time spent in communication routines is the same as when these routines
are called back-to-back with no computation
% :
(i.e., these communications are not overlapped).
We can see on Figure \ref{fig:exp:p2p} that the time is actually
% spend
spent in \verb|shmem_quiet|, which is completing
% (and
(actually performing) the communication.

\ignore{
We are showing the ratio between the time to complete a non-blocking communication ({\tt shmem\_put\_nbi} or {\tt shmem\_get\_nbi} followed by {\tt shmem\_quiet}) and the time spent in these routines surrounding a call to {\tt sleep}, used to let the communications progress in the background (of course, this {\tt sleep} is excluded from the time measurement).

We can see this ratio on Figure \ref{fig:exp:overlap:2nodes}, and we can see that it remains between 2 and 3\% of the total time. Therefore, this communication library has excellent communication and computation overlap capabilities.
}

\subsection{Collective communications: broadcast}
\label{sec:expe:bcast}

\begin{figure}[H]
\vspace*{-.2in}
\centering
\begin{subfigure}[b]{0.49\textwidth}
\includegraphics[width=\linewidth]{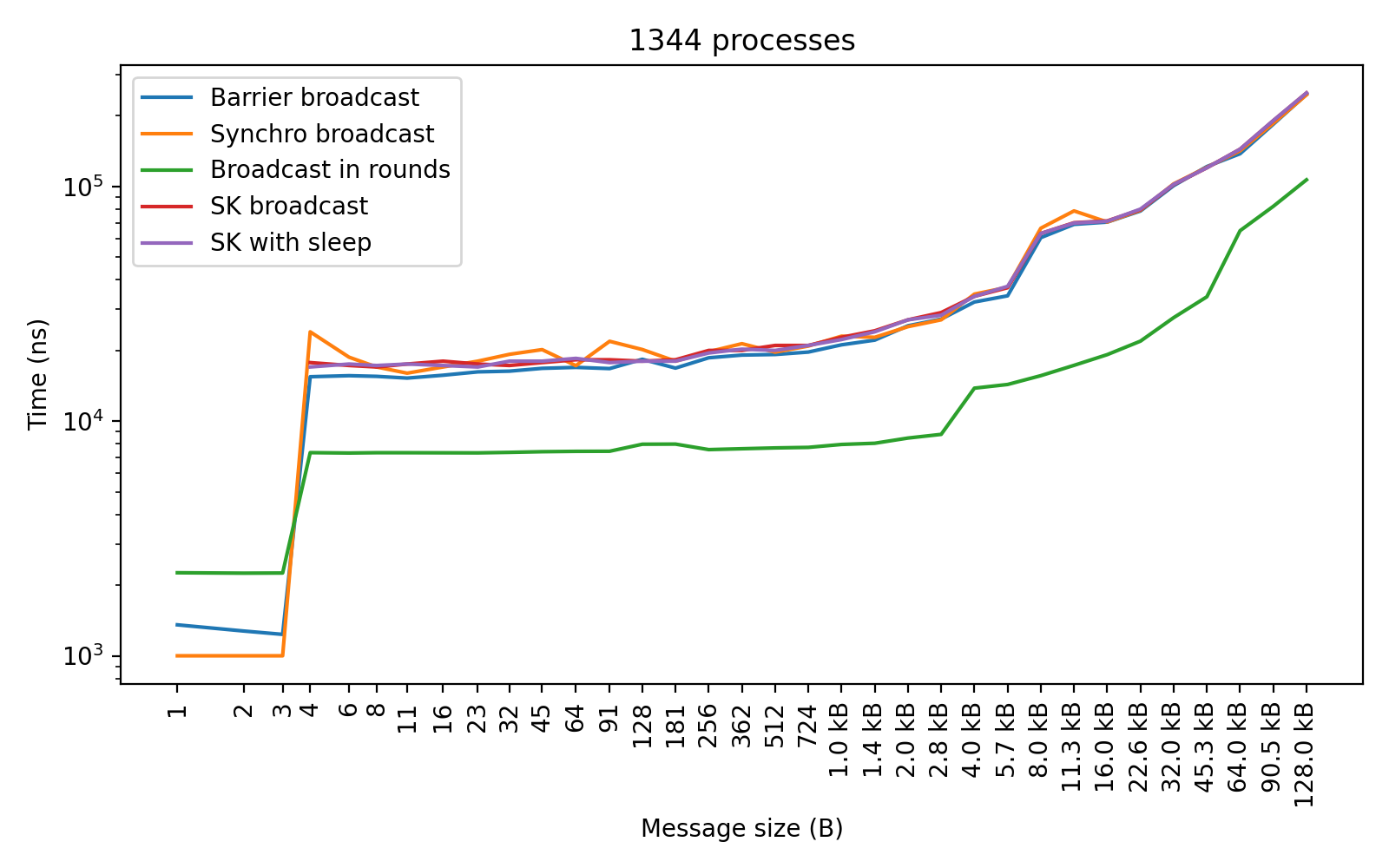}
\caption{\label{fig:exp:bcast:1344}1\,344 processes.}
\end{subfigure}
\hfill
\begin{subfigure}[b]{0.49\textwidth}
\includegraphics[width=\linewidth]{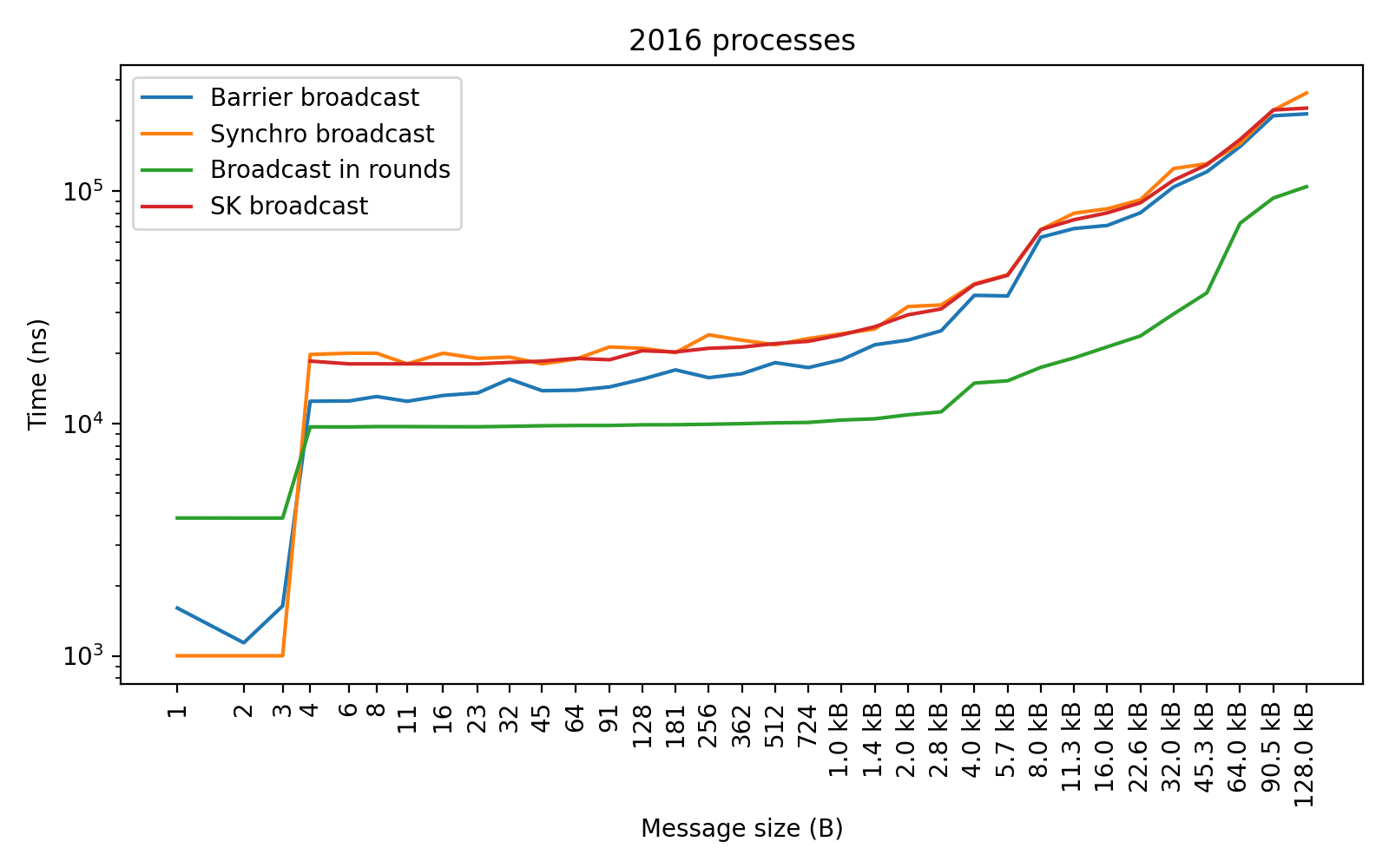}
\caption{\label{fig:exp:bcast:2016}2\,016 processes.}
\end{subfigure}
\caption{\label{fig:exp:bcast}Comparing times returned by the broadcast measurement algorithms.}
%\vspace*{-.2in}
\end{figure}

% We want to make experimental observations on how different from each other the results provided by the various measurement algorithms are.
It is interesting to observe the differences in experimental results for
the various measurement algorithms.
In particular, we expect the \emph{in round} approach to give a smaller value, since consecutive broadcasts can establish a pipeline depending on the communication topology used by the broadcast. 
We
% can
also see on Figure \ref{fig:exp:bcast} that the broadcast separated by barriers can
% also
give smaller measurement values.
As explained in section \ref{sec:expe:bcast},
we
% expected
expect this observation
% that
can be explained
% that
by the final barrier
% might be
being faster on the last processes to finish (see Figure \ref{fig:measure:coll:bcastbarrier}). 

The other algorithms give very similar results. As discussed in section \ref{sec:measure:coll}, the synchronized broadcast can be less reliable because of time drifts and
less significant statistically speaking,
since we are measuring broadcasts one by one.
Moreover, the SK algorithm with consecutive broadcasts separated by a sleep is likely to be more relevant, although here it gives similar results.

\subsection{Locks}
\label{sec:expe:locks}

\begin{figure}[H]
\vspace*{-.2in}
\centering
\begin{subfigure}[b]{0.49\textwidth}
\includegraphics[width=\linewidth]{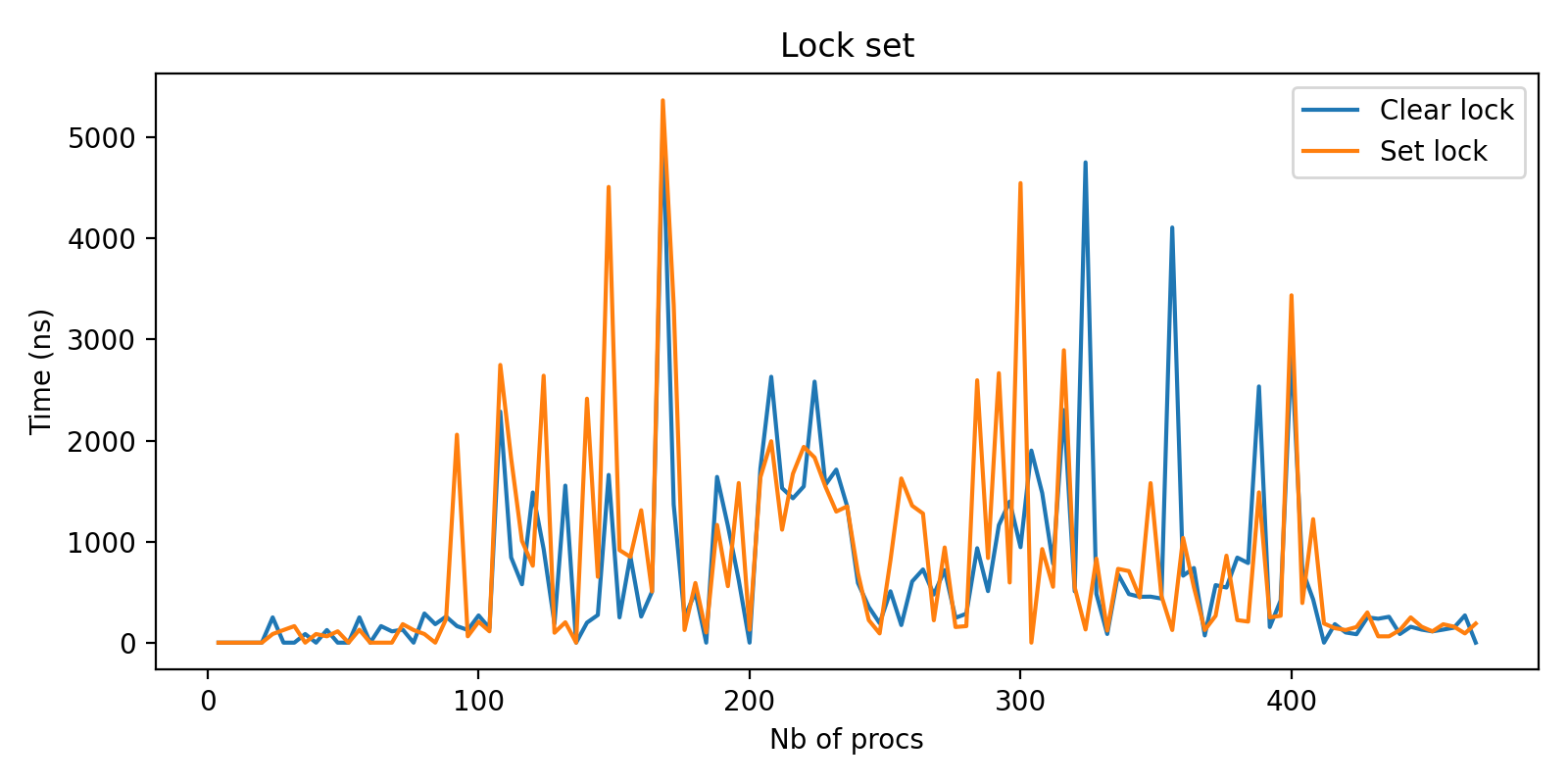}
\caption{\label{fig:exp:locks:set}Acquisition and release.}
\end{subfigure}
\hfill
\begin{subfigure}[b]{0.49\textwidth}
\includegraphics[width=\linewidth]{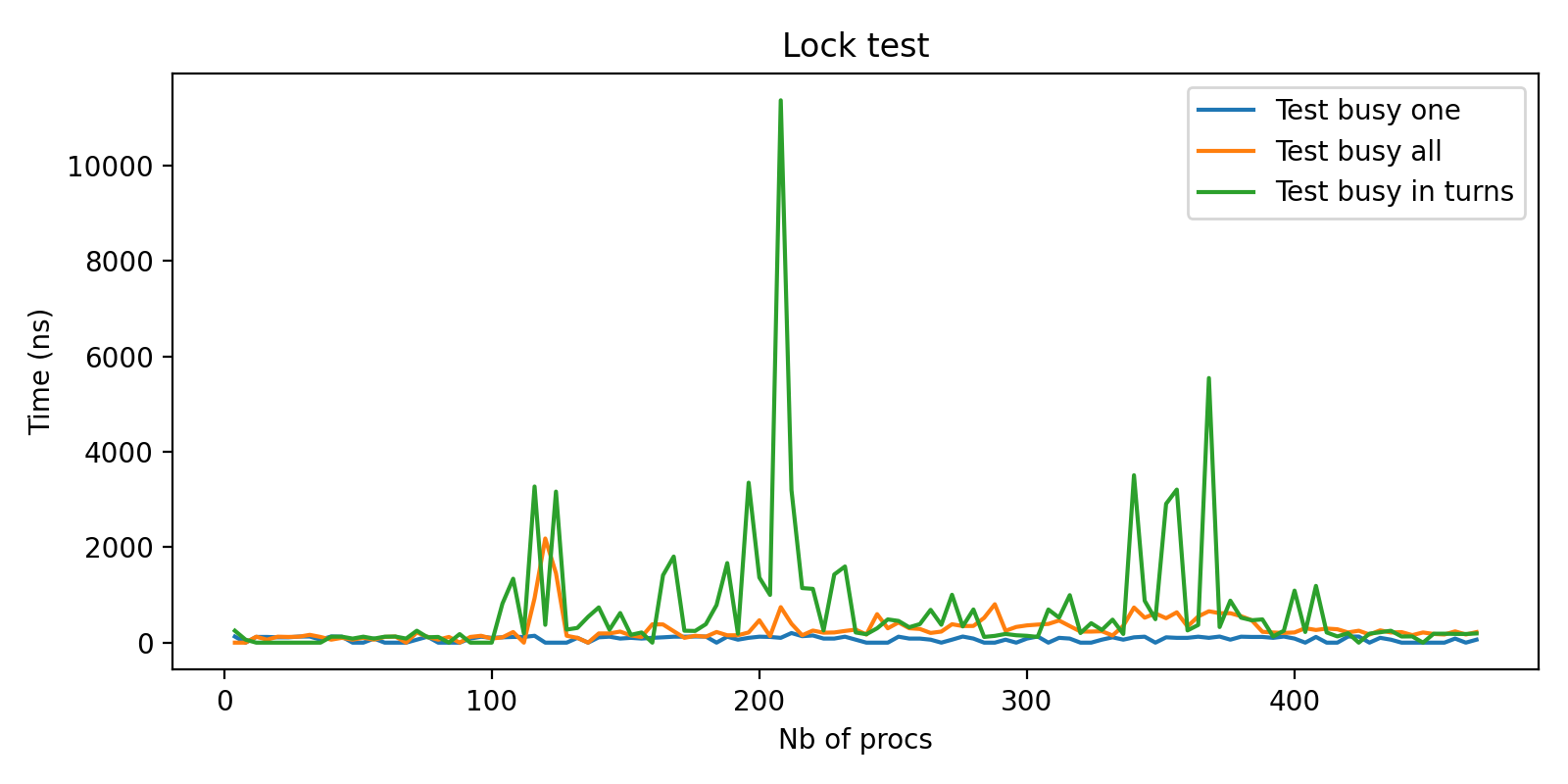}
\caption{\label{fig:exp:locks:test}Test.}
\end{subfigure}
\caption{\label{fig:exp:lock}Global lock functions.}
%\vspace*{-.2in}
\end{figure}

OpenSHMEM provide global locking functions.
% We provide functions to measure the time
Our measurement of the time taken by these functions
% , depending on
takes into account whether the lock is already taken,
% on
who is requesting it, and so on.
Some of these measurements and their scalability
% is
are shown Figure \ref{fig:exp:lock}. 

%% file: conclu.tex
\section{Conclusion and perspectives}
\label{sec:conclu}

Our research work delivers a portable benchmarking framework for
OpenSHMEM, in the spirit of the successful SKaMPI benchmarking system
for MPI.  While the communication libraries are distinct
from one another, the benchmarking methodology practiced in SKaMPI
is more general and very relevant to our OpenSHMEM
benchmarking objectives.  Indeed, we made the important decision
to work within the SKaMPI benchmarking infrastructure and
implement OpenSHMEM-specific functionality, thereby delivering a more
robust outcome in the end.
Clearly, the most important contribution of our research
are the algorithms we created for the unique requirements of
measuring OpenSHMEM routines.

The original SKaMPI benchmarking offered portability across platforms
and the ability to help tune communication library implementation.
Our OpenSHMEM benchmarking development carries forward these
% important
key
% factors
attributes.  To illustrate its use, we conducted experimental evaluation
on the Summit machine.  Our results demonstrate the richness of
performance insight we can gain on point-to-point and collective
operations.  We show how this can be used to optimize certain
implementation parameters.

The increasing complexity of HPC environments will further complicate
abilities to measure their performance.  A well-defined benchmarking
methodology can serve as the core for evaluating multiple
communication libraries.  That perspective is well-supported
based on our experience.  It is reasonable to expect that the
approach we followed of specializing SKaMPI's infrastructure for
OpenSHMEM would work well with other communication models and libraries.

SKaMPI-OpenSHMEM can be downloaded from GitHub at the following address: \url{https://github.com/coti/SKaMPI}.